\documentclass[twocolumn,english,superscriptaddress,showpacs]{revtex4-2}
\usepackage[T1]{fontenc}
\usepackage[utf8]{inputenc}
\usepackage{color}
\usepackage{babel}
\usepackage{bm}
\usepackage{amsmath}
\usepackage{amssymb}
\usepackage{graphicx}
\usepackage[unicode=true,pdfusetitle,
 bookmarks=true,bookmarksnumbered=true,bookmarksopen=true,bookmarksopenlevel=1,
 breaklinks=false,pdfborder={0 0 0},pdfborderstyle={},backref=false,colorlinks=false]
 {hyperref}

\makeatletter
\usepackage{braket}
\usepackage{bm}
\usepackage{tikz}
\usepackage{pgfplots}
\usepackage{pgf}
\usepackage{subfigure}
\usepackage{nicematrix}
\usepackage{diagbox}

\makeatother

\begin{document}
\title{\textcolor{black}{Qu}antum anomalous Hall insulator of composite fermions
in twisted bilayer graphene}
\author{Guangyue Ji}
\affiliation{International Center for Quantum Materials, Peking University, Beijing
100871, China}
\author{Junren Shi}
\email{junrenshi@pku.edu.cn}

\affiliation{International Center for Quantum Materials, Peking University, Beijing
100871, China}
\affiliation{Collaborative Innovation Center of Quantum Matter, Beijing 100871,
China}
\begin{abstract}
\textcolor{black}{We theoretically study the realization of quantum
anomalous Hall insulator (QAHI) of composite fermions (CFs) in the
twisted bilayer graphene (TBG) system. We show that the moiré pattern
in TBG is not only able to provide a commensurate moiré superlattice,
but also a tunable effective periodic potential necessary for the
realization, without the need of imposing an additional superstructure
as in the conventional GaAs system. These make the TBG an ideal platform
for realizing the QAHI of CFs. We establish the phase diagram with
respect to tunable experimental parameters based on the Dirac CF theory.
We find that the topological property of the system depends critically
on the orbital magnetic susceptibility of CFs, which is not specified
in the pristine Dirac CF theory. The experimental realization of QAHI
of CFs would be helpful for unveiling the magnetic property of CFs
and clarifying the issue.}
\end{abstract}
\maketitle

\paragraph*{Introduction}

The moir$\acute{\text{e}}$ pattern in van der Waals (vdW) heterostructures
has attracted widespread investigations in both theoretical and experimental
condensed matter physics. It provides us a powerful way of precisely
controlling the electronic properties~\citep{lopesdossantos_graphene_2007,trambly_de_laissardiere_localization_2010,bistritzer_moire_2011,carr_electronic-structure_2020}
and realizing exotic quantum states including the unconventional superconductor~\citep{cao_unconventional_2018},
Mott insulator~\citep{cao_correlated_2018,chen_evidence_2019}, and
fractional Chern insulator~\citep{xie_fractional_2021}. Moreover,
in the presence of a magnetic\textcolor{black}{{} field, there is an
interplay between the moir$\acute{\text{e}}$ potential and the magnetic
length, and people have used this to study the quantum Hall physics
subjected to a periodic modulation. In contrast to previous works
mostly focused on the study of energy spectrums of non-interacting
electrons, i.e., the Hofstadter’s butterfly~\citep{bistritzer_butterfly_2011,dean_hofstadters_2013,hunt_massive_2013,ponomarenko_cloning_2013},
this work focuses on the quantum anomalous Hall insulator (QAHI) of
composite fermions (CFs), which is a strongly-correlated topological
state consisting of emergent quasiparticles in a Landau level~\citep{zhang_quantum_2014}.}

A CF consists of an electron and $2p$ quantum vortices, and subjects
to a reduced effective magnetic field~\citep{jain_composite-fermion_1989}.
In particular, at even-denominator magnetic filling $\nu_{\text{m}}=1/2p$,
the effective magnetic field vanishes, and the system is proposed
to form a Fermi liquid of CFs (CFL)~\citep{halperin_theory_1993}.
The CFL state has alrea\textcolor{black}{dy been observed in various
2D electron systems including the graphene and its vdW heterostructure~\citep{li_even-denominator_2017,zibrov_tunable_2017}.
Moreover, inspired by the Haldane model~\citep{haldane_model_1988},
it is proposed that imposing a proper electrostatic potential modulation
on the CFL state, which is equivalent to an effective magnetic field
modulation for CFs~\citep{halperin_theory_1993,jain2007composite},
is able to induce the QAHI of CFs~\citep{zhang_quantum_2014}. The
exotic state is proposed to be able to exhibit fractional quantum
Hall effect (FQHE) when $p>1$. As a topological phase consisting
of emergent quasiparticles in a strongly-correlated system, which
is also proposed to lie in a duality web in $2+1$ dimension$\,$\citep{seiberg_duality_2016,senthil_duality_2019},
its novel properties are urged to be tested in experiment. }

\textcolor{black}{However, although the proposal of QAHI of CFs has
attracted much attention, it has yet to be realized in experiment
as far as we know, let alone test its novel physical properties. The
reason is that realizing the QAHI of CFs requires the length scale
of the modulation potential to be commensurate with the magnetic length,
which is difficult to realize in the conventional GaAs system. Besides,
imposing a superstructure on the GaAs sample to construct the required
modulation potential will inevitably affect the quality of the sample.
These difficulties hinder the experimental realization of the QAHI
of CF in the conventional GaAs system. }

\textcolor{black}{In this work, we will focus on its experimental
realization in a moir$\acute{\text{e}}$ heterostructure, i.e., the
twisted bilayer graphene (TBG) system instead. We show that the moiré
pattern in TBG is not only able to provide a commensurate moiré superlattice,
but also a tunable effective periodic potential necessary for the
realization, without the need of imposing an additional superstructure
as in the conventional GaAs system. These make the TBG an ideal platform
for realizing the QAHI of CFs. Furthermore, we establish the phase
diagram with respect to tunable experimental parameters based on the
Dirac CF theory. We find that the topological property of the system
depends critically on the orbital magnetic susceptibility of CFs,
which is not specified in the pristine Dirac CF theory~\citep{son_is_2015}.
The experimental realization of QAHI of CFs will be helpful for unveiling
the magnetic property of CFs and classifying the issue.}

\paragraph*{TBG system}

\textcolor{black}{The QAHI of CFs was first proposed in Ref.~\citep{zhang_quantum_2014}.
According to the CF theory, we know CFs feel a reduced effective magnetic
field: $B^{*}=B-2pn_{\text{e}}\Phi_{0}$, where $B$ is the external
magnetic filed, $n_{\text{e}}$ is the electron density, $\Phi_{0}=h/e$
is a quantum flux. At even-denominator magnetic filling $\nu_{m}\equiv n_{\text{e}}h/eB=1/2p$,
CFs feel zero effective magnetic filed and form the CFL state. Furthermore,
if one imposes a periodic scalar potential modulation on the system,
it will induce an electron density modulation $\delta n_{\text{e}}$,
which will further induce an effective magnetic field modulation $\delta B^{*}$.
By tuning $\delta B^{*}$ in analogy to the Haldane model, it is expected
to realize the QAHI of CFs when integer CF Bloch bands are fully filled.
The band filling factor is defined as $\nu_{\text{e}}\equiv n_{\text{e}}S$,
where $S=3\sqrt{3}L^{2}/2$ is the area of the hexagonal unit cell
and $L$ is the lattice constant. Ref.~\citep{zhang_quantum_2014}
shows that there are two key conditions in realizing the QAHI of CFs.
One is that the lattice constant $L$ should be commensurate with
the magnetic length $l_{B}=\sqrt{\hbar/eB}$. Concretely, it requires
$L=\sqrt{4\pi\nu_{\text{e}}/3\sqrt{3}\nu_{m}}l_{B}$. The other is
a proper periodic modulation, which should be able to fully open the
gap at the Fermi surface and drive the system into an insulator.}

\begin{figure}[t]
\includegraphics[width=1\columnwidth]{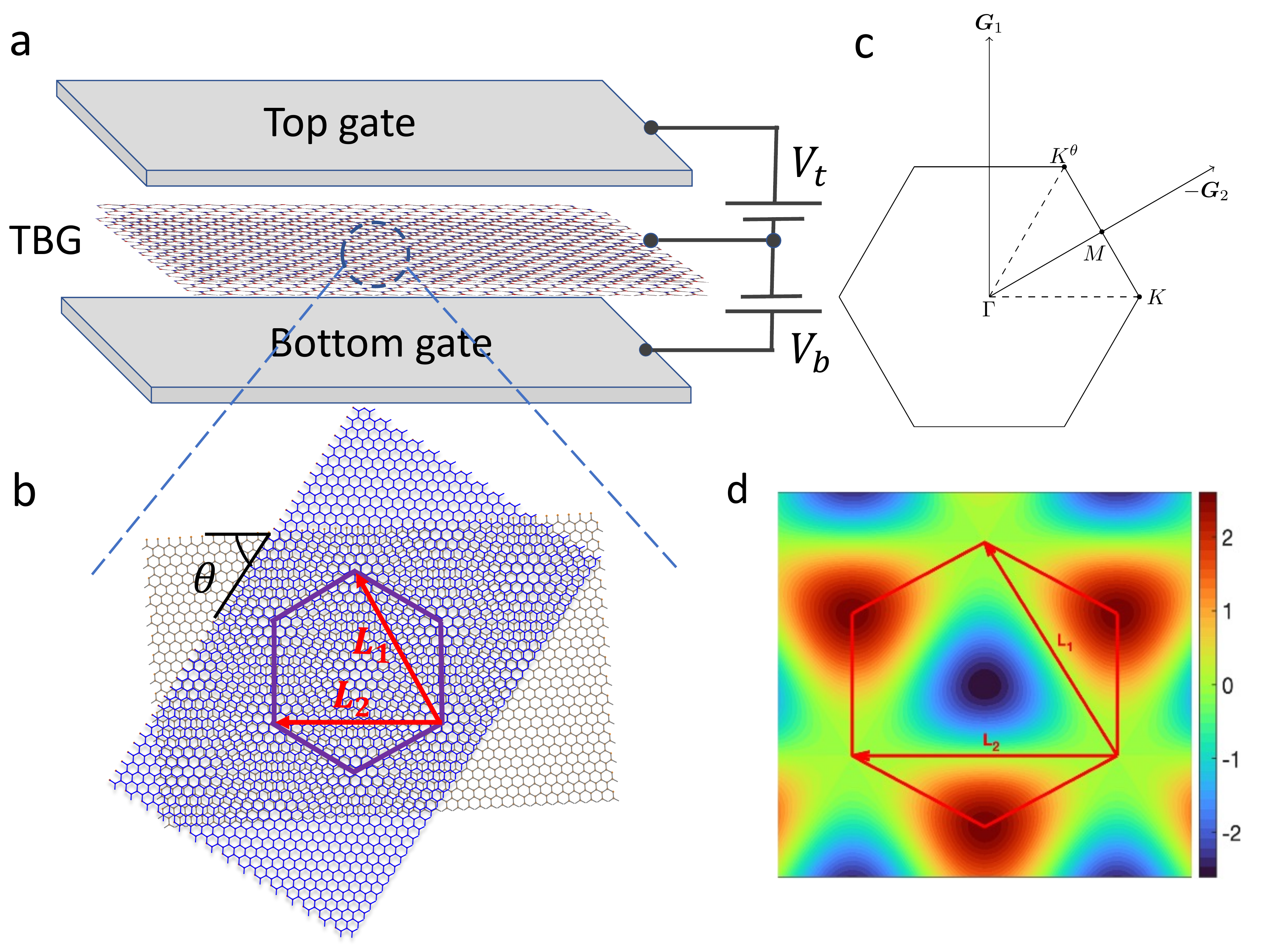}\caption{\label{setup}\textcolor{black}{(a) Schematic of the device with the
TBG encapsulated between two dual-gates. By tuning $V_{t}$ and $V_{b}$,
one can independently tune the electron density $n_{\text{e}}=c_{t}V_{t}+c_{b}V_{b}$
and the electrical displacement field $D=(c_{t}V_{t}-c_{b}V_{b})/2$,
where $c_{t}$ ($c_{b}$) is the unit-area capacitance of the top
(bottom) gate. (b) Moir$\acute{\text{e}}$ Superlattice of TBG. $\bm{L}_{1}=(\sqrt{3}L/2)(-1,\sqrt{3})$
and $\bm{L}_{2}=-\sqrt{3}L(1,0)$ are its primitive lattice vectors.
The lattice constant $L$ can be tuned with the rotation angle $\theta$.
(c) First Brillouin zone. $\bm{G}_{1}$ and $\bm{G}_{2}$ are the
reciprocal lattice vectors of the moiré superlattice. $K$ ($K^{\theta}$)
is the (rotated) Dirac point. (d) Spatial profiles of the periodic
potential $V(\bm{r})$ in Eq.$\,$(\ref{eq:effective_potential_2})
with $\vert V_{0}\vert=1$ and $\nu_{\text{m}}/\nu_{\text{e}}=1/2$.}}
\end{figure}

\textcolor{black}{The required conditions can be fulfilled in TBG
system with the set-up shown in Fig.$\,$\ref{setup}(a), where the
TBG sample is encapsulated with two dual-gates$\,$\citep{zhang_direct_2009}.
This set-up has been widely used to study the properties of bilayer
graphene$\,$\citep{zhang_direct_2009,li_even-denominator_2017,spanton_observation_2018,polshyn_topological_2021}.
By tuning the voltages applied to the top and bottom gates $V_{t}$
and $V_{b}$, one can independently tune the total electron density
$n_{\text{e}}$ and electrical displacement field $D$. Concretely,
the electron density is given by $n_{\text{e}}=c_{t}V_{t}+c_{b}V_{b}$,
where $c_{t}=\varepsilon_{t}/d_{t}$ ($c_{b}=\varepsilon_{b}/d_{b}$)
is the unit-area capacitance of the top (bottom) gate and it depends
on the dielectric constant and thickness of the dielectric layer between
TBG and the top (bottom) gate. The electrical displacement field is
given by $D=(c_{t}V_{t}-c_{b}V_{b})/2$, which can be tuned to be
as large as several volts per nanometer in experiment$\,$\citep{zhang_direct_2009}.
The electrical displacement field will further induce an interlayer
bias $\Gamma$. For the AB-stacked bilayer graphene at the charge
neutral point, a large band gap $\Delta\approx\Gamma\approx0.25\,\text{eV}$
induced by the interlayer bias has already been observed in experiment$\,$\citep{zhang_direct_2009}.
This indicates the interlayer bias $\Gamma$ is widely tunable. In
the following, we will show that the two key conditions for realizing
the QAHI of CFs can be fulfilled with the continuously tunable electron
density $n_{\text{e}}$ and interlayer bias $\Gamma$. }

\textcolor{black}{First, let's look at the commensurate condition.
For a typical magnetic field $B\sim10\,\text{T}$, $L\sim18\,\text{nm}$
for $\nu_{m}=1/2$ and $\nu_{\text{e}}=1$. The ratio between $L$
and the carbon-carbon distance of graphene $a_{0}\approx1.42\,\mathring{\text{A}}$
is $L/a_{0}\sim126$, which corresponds to a large superlattice. On
the other hand, we know that the lattice constant of a TBG superlattice
is determined by its twist angle $\theta$: $L=a_{0}\left|m-n\right|/2\sin(\theta/2)$~\citep{moon_energy_2012},
where the superlattice vector is set to be $\bm{L}_{1}=m\bm{a}_{1}+n\bm{a}_{2}$
and $\bm{a}_{i}$ are the lattice vectors of the graphene. We have
two ways to construct a large superlattice: one is large $m$ and
$n$ and small $\left|m-n\right|$ with $\theta\sim0$, and the other
is large $\left|m-n\right|$ with $\theta\sim30^{\circ}$~\citep{trambly_de_laissardiere_localization_2010}.
By tuning the twist angle $\theta$, one can easily realize a large
superlattice as shown in Fig.$\,$\ref{setup}(b). Furthermore, by
tuning the electron density $n_{\text{e}}$, one can realize the desired
electron filling $\nu_{e}$. Meanwhile, one can tune the applied
magnetic field and realize the desired magnetic filling $\nu_{m}$.
In this way, one can precisely realize the required commensurate condition
in this system.}

\textcolor{black}{Next, let's look at the periodic modulation condition.
In the following, we will show that apart from providing us a moir$\acute{\text{e}}$
superlattice, the moir$\acute{\text{e}}$ pattern is also able to
induce an effective period potential with tunable strength. The TBG
consists of two layers of graphene with a relative rotation angle.
The monolayer graphene is a honeycomb lattice of carbon atoms, of
which each unit cell contains two inequivalent atoms denoted as $\textrm{A}$
and $\textrm{B}$. In the reciprocal lattice space near the Dirac
point $\bm{K}=(4\pi/3\sqrt{3}a_{0})(1,0)$, it can be described by
the Dirac equation $H_{\text{G}}=-iv_{\text{e}}\bm{\sigma}\cdot\bm{\nabla}$,
where $\bm{\sigma}=(\sigma_{x},\sigma_{y})$ are Pauli matrices and
$v_{\text{e}}\approx1\times10^{6}\,\text{m/s}$ is the Fermi velocity~\citep{castro_neto_electronic_2009}.
For the other layer with rotation $\theta$, its Hamiltonian is $H_{\text{G}}^{\theta}=\bm{\sigma}^{\theta}\cdot(-iv_{\text{e}}\bm{\nabla}-\Delta\bm{K})$,
where $\bm{\sigma}^{\theta}=e^{i\theta\sigma_{z}/2}(\sigma_{x},\sigma_{y})e^{-i\theta\sigma_{z}/2}$
are the rotated Pauli matrices, $\Delta\bm{K}=\bm{K}^{\theta}-\bm{K}$
is the displacement of the Dirac point (see Fig.$\;$1c), and $\bm{K}^{\theta}=R(\theta)\bm{K}$
with $R(\theta)$ being the rotation matrix. Furthermore considering
the interlayer hopping, we obtain the following Hamiltonian describing
the TBG~\citep{lopesdossantos_graphene_2007,bistritzer_moire_2011,bistritzer_butterfly_2011}:
\begin{align}
H_{\text{TBG}} & =\left(\begin{array}{cc}
-i\bm{\sigma}\cdot\bm{\nabla} & T(\bm{r})\\
\left[T(\bm{r})\right]^{\dagger} & \bm{\sigma}^{\theta}\cdot(-i\bm{\nabla}-\Delta\bm{K})
\end{array}\right),
\end{align}
where $T(\bm{r})=\omega\sum_{\bm{G}}T_{\bm{G}}e^{i\bm{G}\cdot\bm{r}}$
is the interlayer hopping matrix with 
\begin{gather}
T_{0}=\left(\begin{array}{cc}
1 & 1\\
1 & 1
\end{array}\right),T_{-\bm{G}_{1}}=\left(\begin{array}{cc}
e^{i\phi_{0}} & 1\\
e^{-i\phi_{0}} & e^{i\phi_{0}}
\end{array}\right),\nonumber \\
T_{-\bm{G}_{1}-\bm{G}_{2}}=\left(\begin{array}{cc}
e^{-i\phi_{0}} & 1\\
e^{i\phi_{0}} & e^{-i\phi_{0}}
\end{array}\right)\label{eq:hopping matrix}
\end{gather}
for the initial $\textrm{AB}$ stacked structure, $\bm{G}_{1}=\sqrt{3}R(-\pi/6)\Delta\bm{K}$
and $\bm{G}_{2}=\sqrt{3}R(\pi/2)\Delta\bm{K}$ are the reciprocal
lattice vectors of the moiré superlattice, $\omega\approx0.11\,\text{eV}$
is the hopping strength and $\phi_{0}=2\pi/3$~\citep{lopesdossantos_graphene_2007}.}

\textcolor{black}{In the presence of a perpendicular magnetic field,
the system will form Landau levels. For a monolayer graphene with
a linear dispersion, its LL spectrum is $E_{n}=\text{sgn}(n)\sqrt{2\vert n\vert eB\hbar}v_{\text{e}}$
with $n=0,\pm1,\ldots$. In this work, we focus on the spin and valley-polarized
zeroth LL (ZLL), which is reasonable in experiment with strong magnetic
field$\,$\citep{li_even-denominator_2017} and electric displacement
field$\,$\citep{huang_valley_2022}. The wavefunctions in the ZLL
of layer $1$ have only $\textrm{A}$ atom component, and we denote
them as $\vert\psi_{1,m}\rangle=(\vert m\rangle,0)^{T}$, where $\vert m\rangle$
are the LLL eigenstates of the conventional massive electron system.
The ZLL eigenstates of layer $2$ are related to layer $1$ and given
by $\vert\psi_{2,m}\rangle=e^{i\theta\sigma_{z}/2}e^{i\Delta\bm{K}\cdot\bm{r}}\vert\psi_{1,m}\rangle$.
In the presence of the interlayer coupling $T(\bm{r})$, the ZLLs
of the two layers will couple with each other. In the following, we
will show that the interlayer coupling will induce an effective periodic
potential.}

\textcolor{black}{In the presence of non-zero interlayer bias $\Gamma$,
the layer ZLL energy degeneracy will be lifted. When $\Gamma$ is
much larger than the interlayer hopping strength $\omega$, the two
layers can be well seen as two isolated layers with the interlayer
coupling acting as an effective perturbation potential. The effective
potential induced by $T(\bm{r})$ can be derived by employing the
perturbation theory. Its matrix element (accurate to the second order
of $\omega$) between any two states $\psi_{1,m}$ and $\psi_{1,m^{\prime}}$
in the ZLL of layer $1$ is
\begin{align}
V_{mm^{\prime}} & =\frac{\omega^{2}}{\Gamma}\sum_{l,\bm{G},\bm{G}^{\prime}}T_{\bm{G}}^{11}(T_{\bm{G}^{\prime}}^{11})^{*}\langle m\vert e^{i\bm{q}\cdot\bm{r}}\vert l\rangle\langle l\vert e^{-i\bm{q}^{\prime}\cdot\bm{r}}\vert m^{\prime}\rangle,\label{eq:effective_potential}
\end{align}
where $\bm{q}=\bm{G}+\Delta\bm{K}$. Note that $\sum_{l}\vert l\rangle\langle l\vert$
is in fact the conventional LLL projection operator $\hat{P}_{\text{LLL}}$$\:$\citep{girvin_formalism_1984}.
By using the commutation relation of projected plane wave factors$\:$\citep{girvin_formalism_1984},
one can show $\sum_{l}\langle m\vert e^{i\bm{q}\cdot\bm{r}}\vert l\rangle\langle l\vert e^{-i\bm{q}^{\prime}\cdot\bm{r}}\vert m^{\prime}\rangle=e^{-\frac{1}{2}q^{\prime}\bar{q}}\langle m\vert e^{i(\bm{q}-\bm{q}^{\prime})\cdot\bm{r}}\vert m^{\prime}\rangle$,
where $q=q_{x}+iq_{y}$ and $\bar{q}=q_{x}-iq_{y}$. By substituting
it and the hopping matrix Eq.$\,$(\ref{eq:hopping matrix}) into
Eq.$\,$(\ref{eq:effective_potential}), we derive $V_{mm^{\prime}}=\langle m\vert V(\bm{r})\vert m^{\prime}\rangle$
with the effective potential
\begin{align}
V(\bm{r}) & =V_{0}\left(\sum_{i=1}^{2}e^{i\bm{G}_{i}\cdot\bm{r}}+e^{-i(\bm{G}_{1}+\bm{G}_{2})\cdot\bm{r}}\right)+\text{h.c.},\label{eq:effective_potential_2}
\end{align}
where $V_{0}=\frac{\omega^{2}}{\Gamma}\exp\left[-(\vert q\vert^{2}l_{B}^{2}/2)e^{i2\pi/3}\right]$,
$\vert q\vert=\vert\Delta\bm{K}\vert=2\vert\bm{K}\vert\sin(\theta/2)$,
and an irrelevant phase factor $e^{-i\phi_{0}}$ has been removed
through a translation of the original point$\:$\citep{zhang_topological_2020}.
Furthermore, by using the commensurate condition, it can be simplified
to be
\begin{equation}
V_{0}=\frac{\omega^{2}}{\Gamma}\exp\left[\alpha/\sqrt{3}\right]\exp\left[-i\alpha\right],
\end{equation}
where $\alpha=\frac{\pi}{3}\frac{\nu_{\text{m}}}{\nu_{\text{e}}}$
is the phase factor and it is determined by the ratio of magnetic
and band filling factors. The spatial profile of $V(\bm{r})$ with
$\vert V_{0}\vert=1$ and $\nu_{\text{m}}/\nu_{\text{e}}=1/2$ is
plotted in Fig.$\;$\ref{setup}(d), and it is triangular. Since the
interlayer bias $\Gamma$ is widely tunable, the effective potential
$V_{0}$ induced by the moi\'{r}e pattern is also widely tunable.
Thus, it is expected to fulfill the required periodic modulation condition
without much difficulty.}

\paragraph*{Dirac composite fermions}

\textcolor{black}{In the following, we study the QAHI of CFs subjected
to the periodic potential Eq.$\,$(\ref{eq:effective_potential_2})
by adopting the Dirac CF theory, in which the particle-hole symmetry
is explicit~\citep{son_is_2015}. The Dirac CF theory treats CFs
as neutral massless Dirac particles. The motion of Dirac CFs is governed
by the Dirac equation $i\partial_{t}\psi=\hat{H}\psi$, where $\psi$
is the Dirac CF spinor, and the Hamiltonian}
\begin{align}
\hat{H} & =\bm{\sigma}\cdot(-i\bm{\partial}-\bm{a}(\bm{r}))+a_{0}(\bm{r})\sigma_{0},\label{eq:DiracEq}
\end{align}
where $a^{\mu}=(a^{0},\bm{a})$ are the internal gauge fields, $\sigma_{0}$
is the identity matrix, we have set $e=\hbar=v_{\text{CF}}=1$ and
will restore these constants when appropriate. Meanwhile, the CF system
subjects to the constraints~\citep{son_is_2015}
\begin{align}
\rho_{\textrm{CF}} & \equiv\left\langle \psi^{\dagger}\psi\right\rangle =\frac{B}{4\pi},\label{eq:CFDensity}\\
\bm{j}_{\textrm{CF}} & \equiv\left\langle \psi^{\dagger}\bm{\sigma}\psi\right\rangle =-\frac{1}{4\pi}\hat{z}\times\bm{E},\label{eq:CFCurrent}
\end{align}
where $\rho_{\textrm{CF}}$ and $\bm{j}_{\textrm{CF}}$ are the CF
density and current, respectively, $\bm{B}=B\hat{z}$ and $\bm{E}$
are the external electromagnetic fields, and $\left\langle \cdots\right\rangle $
denotes the grand canonical ensemble average. The CF density is set
by the external magnetic field, but not the electron density as assumed
in the HLR theory~\citep{halperin_theory_1993}. It is a manifestation
of the particle--vortex duality with the electron system. In the
presence o\textcolor{black}{f the scalar potential $V(\bm{r})$, the
CF current is fixed to be $\bm{j}_{\textrm{CF}}=(1/4\pi)\hat{z}\times\bm{\nabla}V(\bm{r})$
according to Eq.~(\ref{eq:CFCurrent}). To induce a CF current satisfying
this constraint, the internal magnetic field $\bm{b}(\bm{r})$ coupled
to CFs needs to adjust accordingly. By tuning it properly, it is expected
to realize the QAHI of Dirac CFs.}

\textcolor{black}{Different from systems governed by the Schr}\textcolor{black}{\"{o}}\textcolor{black}{dinger
equation, there exists a Dirac sea consisting of all negative-energy
states. When a modulation is imposed on the system, not only positive-energy
states will respond to it, but also the Dirac sea. In Ref.~\citep{principi_linear_2009},
Principi et al. show that the current response of the Dirac sea to
a vector potential is divergent. Thus, to evaluate the physical response
current, we shall regularize the system first.}

\textcolor{black}{We regularize the system by employing the Pauli-Villars
regularization scheme for $(2+1)$-dimensional massless Dirac fermions
proposed by Redlich in Refs.~\citep{redlich_gauge_1984,redlich_parity_1984}.
The regularization scheme is as follows. For a massless Dirac system,
one can introduce two auxiliary massive systems with the Hamiltonian
$\hat{H}_{m}=\hat{H}\pm m\sigma^{3}$, where $m$ is the mass and
its magnitude is required to be much larger than the chemical potential,
i.e., $m\gg\mu$. By canceling out the divergent term in our massless
system with the auxiliary systems, we obtain a finite regularized
current 
\begin{equation}
\bm{j}^{\text{R}}\equiv\bm{j}_{m=0}-(\bm{j}_{m}+\bm{j}_{-m})/2+\nabla\times(\chi_{m}^{\text{mag}}\bm{b}),\label{eq:current_Reg}
\end{equation}
where $\bm{j}_{\pm m}$ are the response currents of the massive systems.
In addition to a divergent term, $\bm{j}_{\pm m}$ also contains finite
Chern-Simons current and magnetization current $\nabla\times(\chi_{m}^{\text{mag}}\bm{b})$,
where $\chi_{m}^{\text{mag}}=1/12\pi m$ is the orbital magnetic susceptibility~\citep{koshino_orbital_2011}.
Since the Chern-Simons current is odd with respect to the mass, we
eliminate it by introducing two auxiliary massive systems with opposite
masses. Because the magnetization current is even with respect to
the mass, we need to subtract its contribution manually as shown in
Eq.~(\ref{eq:current_Reg}). }

\begin{figure}[t]
\includegraphics[width=1\columnwidth]{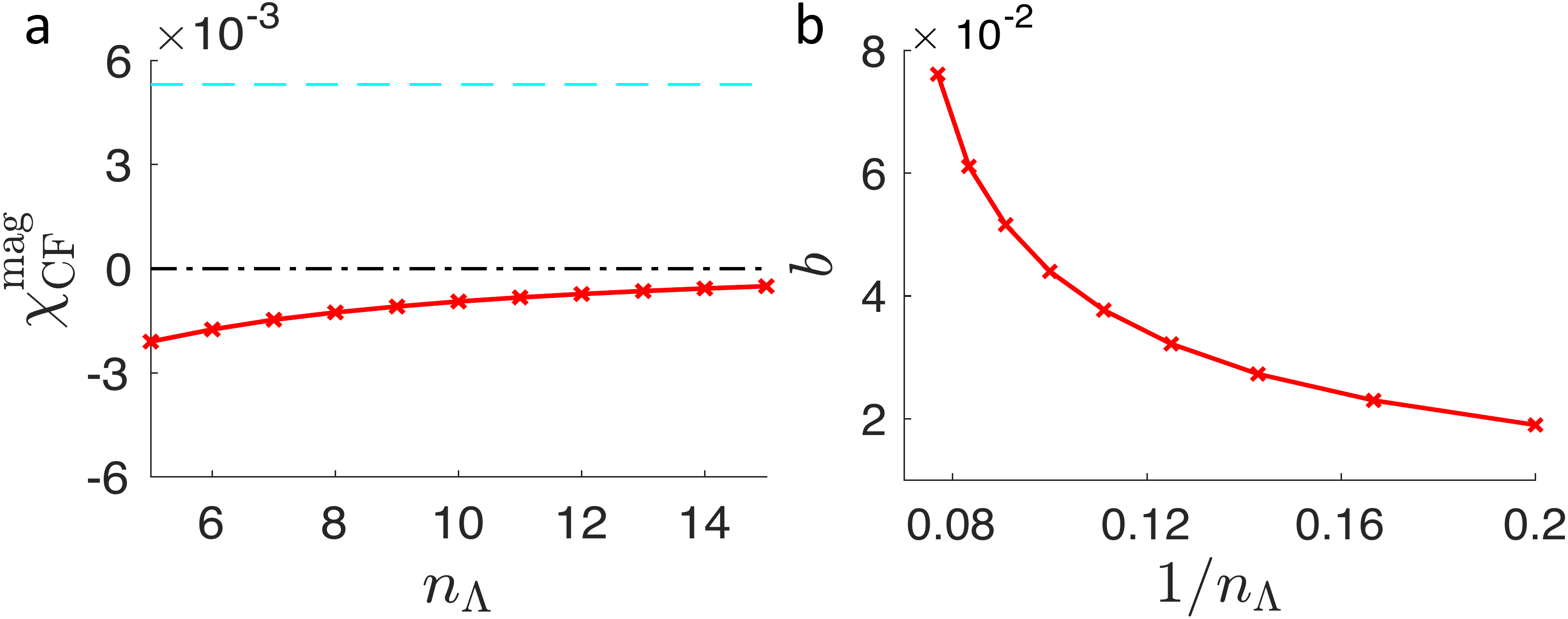}

\caption{\label{magnetic_sus}(a) CF magnetic susceptibility $\chi_{\text{CF}}^{\text{mag}}$
vs momentum cut-off $\Lambda=n_{\Lambda}\vert\bm{G}_{i}\vert$. The
red line denotes the magnetic susceptibility of massless Dirac CFs.
It approaches zero with the increase of the momentum cut-off. The
horizontal cyan dashed line denotes $\chi_{m}^{\text{mag}}=1/12\pi m$
with $m=5$. (b) Induced Chern-Simons field strength (in unit of $\hbar/eL^{2}$)
vs momentum cut-off. The modulation strength $\vert V_{0}\vert$ is
set to be $5\times10^{-3}$. The induced Chern-Simons magnetic field
is divergent with the increase of the momentum cut-off. It results
from the zero CF magnetization susceptibility implicitly assumed in
the Dirac CF theory. }
\end{figure}

\textcolor{black}{To check the }correctness of the above regularization
scheme, we numerically calculate the orbital magnetization susceptibility
$\chi_{\text{CF}}^{\text{mag}}$ of our massless Dirac system in the
linear response region. First, we solve the eigenequation $\hat{H}\psi_{n,\bm{k}}(\bm{r})=E_{n}(\bm{k})\psi_{n,\bm{k}}(\bm{r})$
subjected to a given $\bm{b}$ for both massive and massless systems
by employing the plane wave expansion method. Next, we calculate the
current $\bm{j}_{\textrm{m=0}}$ and $\bm{j}_{\textrm{m}}$ with derived
eigenstates, and derive the regularized CF current $\bm{j}^{\text{R}}$
by applying Eq.~(\ref{eq:current_Reg}). Then, we can derive $\chi_{\text{CF}}^{\text{mag}}$
directly. The results show that $\chi_{\text{CF}}^{\text{mag}}\rightarrow0$
with the increase of the momentum cut-off $\Lambda=n_{\Lambda}\vert\bm{G}_{i}\vert$
(see Fig.$\;$\ref{magnetic_sus}(a)). This is consistent with the
analytic result that the orbital magnetic susceptibility of massless
Dirac fermions at $\mu>0$ is zero~\citep{koshino_orbital_2011}.

However, a zero CF magnetization susceptibility will lead to an infinite
electron density response. The density susceptibility of electrons
$\chi_{\text{e}}$ is inversely proportional to the magnetic susceptibility
of CFs $\chi_{\text{CF}}^{\text{mag}}$ as follows~\citep{son_is_2015}
\begin{equation}
\chi_{\text{e}}=\left(\frac{e}{2h}\right)^{2}\frac{1}{\chi_{\text{CF}}^{\text{mag}}}.\label{eq:ChiE_ChiCFMag}
\end{equation}
N\textcolor{black}{ote that though different CF effective theories
treat CFs as different types of particles, the above relation is identical~\citep{son_is_2015,wang_particle-hole_2017,ji_emergence_2021}.
By substituting $\chi_{\text{CF}}^{\text{mag}}=0$ into Eq.~(\ref{eq:ChiE_ChiCFMag}),
one will obtain an infinite density response. That is to say a finite
scalar modulation $V$ will induce an infinite electron density fluctuation,
which is obviously unphysical. In our numerical study, we find that
the induced magnetic filed $\bm{b}$, which is proportional to the
electron density fluctuation given by $b=B-4\pi\rho_{\textrm{e}}$,
is divergent with the increase of the cut-off as shown in Fig.$\;$\ref{magnetic_sus}(b).
This also indicates that the density response of the system is divergent.}

\textcolor{black}{As we can see, the unphysical infinite density response
results from the zero magnetization susceptibility of CFs implicitly
assumed in the Dirac CF theory. In fact, the Dirac theory is a long-wavelength
low-energy effective theory, while the magnetization is an effect
of higher order$\,$\citep{ji_emergence_2021,ji_berry_2020}. In
the following, we proceed by introducing a finite orbital magnetic
susceptibility of CFs $\chi_{\text{CF}}^{\text{mag}}$ into the Dirac
CF theory and treating it as an extra parameter. In this case, there
is an extra orbital magnetization current and the CF current in Eq.~(\ref{eq:CFCurrent})
needs to be modified to be 
\begin{equation}
\bm{j}_{\textrm{CF}}=\left\langle \psi^{\dagger}\bm{\sigma}\psi\right\rangle +\nabla\times(\chi_{\text{CF}}^{\text{mag}}\bm{b}),
\end{equation}
where the $\chi_{\text{CF}}^{\text{mag}}\bm{b}$ is the orbital magnetization.
In this way, we solve the density response divergence problem in the
pristine Dirac CF theory. Furthermore, by carefully tuning the periodic
modulation potential $V(\bm{r})$, we show the system can form the
QAHI of Dirac CFs. Meanwhile, we find that the topological property
of the system depends critically on the }magnetic susceptibility of
CFs.

\paragraph*{QAHI of Dirac CFs}

\begin{figure}[t]
\includegraphics[width=1\columnwidth]{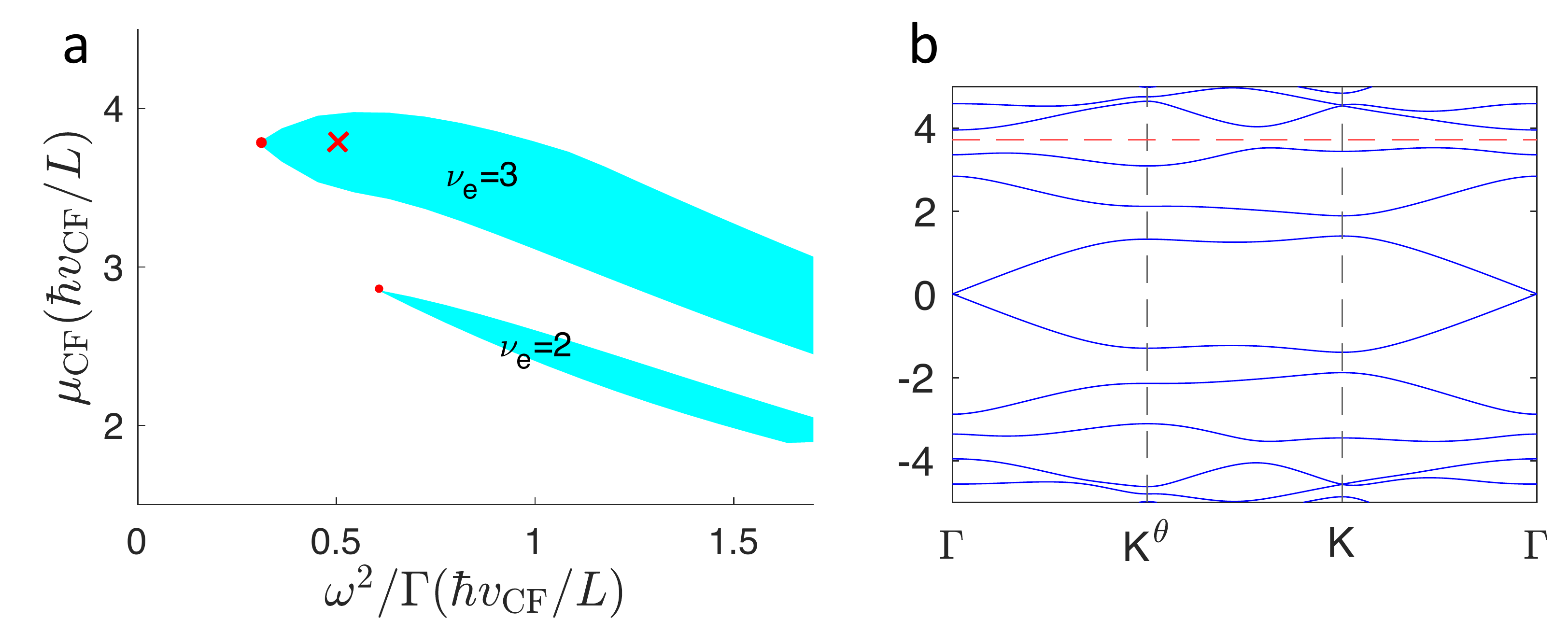}

\caption{\label{Band}(a) \textcolor{black}{Phase diagrams of Dirac CFs for
$\nu_{\text{m}}=1/2$. The area of insulating states is marked with
cyan. $\mu_{\text{CF}}$ is the chemical potential of CFs. The energy
unit is $\hbar v_{\textrm{CF}}/L$. The magnetic susceptibility of
CFs is set to be $\chi_{\text{CF}}^{\text{mag}}=-1/8\pi m_{\text{CF}}$.
The critical bias (marked with red circles) is $\Gamma_{\text{c}}\approx\omega^{2}/(0.6\hbar v_{\textrm{CF}}/L)\approx3.4\epsilon B^{-1/2}\,\text{eV}$
for $\nu_{\text{e}}=2$ and $\Gamma_{\text{c}}\approx8.4\epsilon B^{-1/2}\,\text{eV}$
for $\nu_{\text{e}}=3$. (b) Energy bands of Dirac CFs at $\omega^{2}/\Gamma=0.5$
and $\nu_{\text{e}}=3$ (marked with a cross in (a)). It can seen
that the energy degeneracy near the Fermi surface has been fully lifted
with the given periodic potential modulation. The CF Hall conductivity
of the insulating state is $-e^{2}/2h$ ($+e^{2}/2h$) $\text{for }\Gamma\chi_{\text{CF}}^{\text{mag}}<0$
($\Gamma\chi_{\text{CF}}^{\text{mag}}>0$). The corresponding electron
Hall conductivity is $e^{2}/h$ (0).}}
\end{figure}

\textcolor{black}{In the absence of modulation, CFs are free and form
a Dirac cone with the Fermi wave vector $k_{\text{F}}=\sqrt{eB/\hbar}$.
By folding the linear dispersion in the first BZ, we can obtain the
energy bands of free Dirac CFs. They will touch at points of high
symmetry. In the presence of a periodic modulation $V(\bm{r})$, energy
degeneracies will be partially lifted. With an appropriate modulation
strength and phase, the gap around the Fermi surface can be fully
open, and the system forms an insulating state as shown in Fig.~\ref{Band}(b).
The whole phase diagram for $\nu_{\text{m}}=1/2$ is shown in Fig.~\ref{Band}(a).
Note that the system will not form an insulating state at $\nu_{\text{e}}=1$,
because the degeneracy between the first two positive bands at $K$
point can not be lifted with the triangular potential $V(\bm{r})$
shown in Fig.$\;$\ref{setup}(d). }

\textcolor{black}{In the process of deriving the effective potential
$V(\bm{r})$, it's assumed that the interlayer hopping strength $\omega$
is much smaller than the interlayer bias $\Gamma$ and can be treated
as a perturbation. Now, let's check its rationality. In the case of
$\nu_{\text{m}}=1/2$ and $\nu_{\text{e}}=2$, the ratio between $\omega$
and the critical bias $\Gamma_{\text{c}}$ is $\omega/\Gamma_{\text{c}}\approx0.6\hbar v_{\textrm{CF}}/L\omega$,
where we have set the magnetic susceptibility of CFs $\chi_{\text{CF}}^{\text{mag}}=-1/8\pi m_{\text{CF}}$
as done in Refs.~\citep{wang_particle-hole_2017,ji_emergence_2021}
and $m_{\text{CF}}$ is the effective mass of CFs~\citep{halperin_theory_1993,jain2007composite}.
By substituting the CF velocity $v_{\text{CF}}=\hbar k_{\text{F}}/m_{\text{CF}}\approx(7.2/\epsilon)\times10^{5}\,\text{m/s}$
and the commensurate lattice constant $L=\sqrt{4\pi\nu_{\text{e}}/3\sqrt{3}\nu_{m}}l_{B}$
into the ratio, we obtain $\omega/\Gamma_{\text{c}}\approx(3.2/\epsilon)\times10^{-2}B^{1/2}$,
where $\epsilon$ is the dielectric constant and it is determined
by the substrates. For the typical experimental parameters $\epsilon\sim3-15$
and $B\sim10\,\text{T}$$\,$\citep{toke_su4_2007}, $\omega/\Gamma_{\text{c}}\ll1$.
Thus, there is a wide range of $\Gamma$ satisfying $\omega\ll\Gamma<\Gamma_{\text{c}}$.
Since the interlayer bias $\Gamma$ is widely tunable, e.g., $\Gamma\approx0.25\,\text{eV}>\omega\approx0.11\,\text{eV}$
has already been realized in experiment$\,$\citep{zhang_direct_2009},
it is reasonable to treat the interlayer hopping as a perturbation
to the CFL state.} 

\textcolor{black}{Next, we analyze the topological property of the
insulating state by calculating its Hall conductivity. For a non-degenerate
band, one can evaluate its Hall conductivity by applying the TKNN
formula: $\sigma^{xy}=-\mathcal{C}/2\pi=-(1/2\pi)\int_{\textrm{BZ}}(d^{2}k/2\pi)\Omega_{k_{x}k_{y}}$,
where $\Omega_{k_{x}k_{y}}$ is the Berry curvature in the momentum
space, and $\mathcal{C}$ is the Chern number of the band~\citep{thouless_quantized_1982}.
Naively, one expects to derive the Hall conductivity of the system
by summing up the Chern number of all occupied positive bands. However,
this is incorrect for the Dirac system because it ignores the contribution
of the filled Dirac sea. To correctly evaluate the Hall conductivity
of the Dirac system, we regularize the system by introducing auxiliary
massive systems as done in Eq.~(\ref{eq:current_Reg}). The expression
of the regularized Hall conductance is given by
\begin{equation}
\sigma_{\text{CF}}^{xy}\equiv\sigma_{m=0}^{xy}-(\sigma_{m}^{xy}+\sigma_{-m}^{xy})/2,\label{eq:Hall_conductivity_reg}
\end{equation}
where $\sigma_{m=0}^{xy}$ ($\sigma_{m}^{xy}$) is the total Hall
conductivity of all occupied bands below the chemical potential $\mu_{\text{CF}}$
($\ll m$) of the massless (massive) Dirac system. In numerical calculations,
we evaluate $\sigma_{m}^{xy}$ by setting a momentum cut-off $\Lambda=n_{\Lambda}\vert\bm{G}_{i}\vert$
as done before. Since $\sigma_{\text{CF}}^{xy}$ is a topological
quantity independent of the cut-off, it can be easily evaluated by
setting a small $n_{\Lambda}$. Finally, we obtain the Hall conductivity
of the insulating state, and it is (in unit of $e^{2}/h$)
\begin{equation}
\sigma_{\text{CF}}^{xy}=\begin{cases}
-\frac{1}{2}, & \text{for }\Gamma\chi_{\text{CF}}^{\text{mag}}<0\\
+\frac{1}{2}, & \text{for }\Gamma\chi_{\text{CF}}^{\text{mag}}>0
\end{cases}.\label{eq:CF_Hall}
\end{equation}
The half-integer quantization is a character of Dirac particles~\citep{schakel_relativistic_1991,mikitik_manifestation_1999}.
It can be seen that the Hall conductivity of the system critically
depends on the orbital magnetization susceptibility of CFs. The reason
is that the induced internal magnetic field are opposite for positive
and negative $\chi_{\text{CF}}^{\text{mag}}$.}

\textcolor{black}{With the CF Hall conductivity determined, we can
derive the electron Hall conductivity by applying the conductivity
transformation relation $\sigma_{\text{e}}^{xy}=e^{2}/2h-(e^{2}/2h)^{2}/\sigma_{\text{CF}}^{xy}$~\citep{son_is_2015}.
By substituting the CF conductivity Eq.~(\ref{eq:CF_Hall}) into
the relation, we obtain the electron Hall conductivity (in unit of
$e^{2}/h$)
\begin{equation}
\sigma_{\text{e}}^{xy}=\begin{cases}
1, & \text{for }\Gamma\chi_{\text{CF}}^{\text{mag}}<0\\
0, & \text{for }\Gamma\chi_{\text{CF}}^{\text{mag}}>0
\end{cases}.\label{eq:ele_Hall}
\end{equation}
For $\Gamma\chi_{\text{CF}}^{\text{mag}}<0$, its Hall conductivity
is nonzero and the insulating state is a topological insulator. Thus,
the effective periodic potential induced by the interlayer hopping
of the TBG system can indeed drive the CFL state into the QAHI of
CFs.}

\textcolor{black}{In this work, we focus on the TBG system and the
effective potential given by Eq.$\,$(\ref{eq:effective_potential_2}).
In fact, the phase and strength of the effective potential can be
further tuned by coupling more layers$\,$\citep{lui_observation_2011,bao_stacking-dependent_2011,carr_electronic-structure_2020}.
By using the interference between effective potentials induced by
different layers, it is flexible to realize a desired potential. In
this case, it is also possible to construct a hexagonal periodic potential
and realize the QAHI of CFs at $\nu_{\text{e}}=1$. Besides, one can
also tune the effective potential by other methods, including coupling
to substrates and tuning the strain~\citep{zhou_substrate-induced_2007,hunt_massive_2013,spanton_observation_2018,choi_controlling_2010}.
In short, the moir$\acute{\text{e}}$ pattern in vdW heterostructures
provides us a powerful way of controlling the electronic properties
to realize the QAHI of CFs.}

\textcolor{black}{For the theoretical part, a full understanding of
orbital magnetic susceptibility of CFs is still absent and relevant
investigations are rare. There even exists a misconception that its
effect on observable physical quantities is negligible. However, we
reveal that it in fact plays a central role in the QAHI of CFs. In
particular, the zero magnetic susceptibility of CFs implicitly assumed
in the Dirac theory will lead to an unphysical infinite density response.
Thus, the experimental realization of the QAHI of CFs would be helpful
for unveiling the magnetic property of CFs and clarifying the issue.}

\paragraph*{\textcolor{black}{Summary}}

\textcolor{black}{In summary, we show that the moiré superlattice
and effective periodic potential induced by the the moiré pattern
make TBG an ideal platform for realizing the QAHI of CFs. We establish
the phase diagram with respect to tunable experimental parameters
based on the Dirac CF theory. We find that the topological property
of the system depends critically on the orbital magnetic susceptibility
of CFs, which is not specified in the pristine Dirac CF theory. The
experimental realization of QAHI of CFs would be helpful for clarifying
the issue.}
\begin{acknowledgments}
G.J. and J.S. thank Qian Niu, Di Xiao, Jing-Yuan Chen, Yinhan Zhang,
Haoran Chen and Xuesong H\textcolor{black}{u for valuable discussions.
This work is supported by the National Basic Research Program of China
(973 Program) Grants No. 2018YFA0305603 and No. 2021YFA1401900 and
the National Science Foundation of China Grant No. 12174005.}
\end{acknowledgments}

\bibliographystyle{apsrev4-1}
\bibliography{composite_fermion}

\begin{thebibliography}{45}%
\makeatletter
\providecommand \@ifxundefined [1]{%
 \@ifx{#1\undefined}
}%
\providecommand \@ifnum [1]{%
 \ifnum #1\expandafter \@firstoftwo
 \else \expandafter \@secondoftwo
 \fi
}%
\providecommand \@ifx [1]{%
 \ifx #1\expandafter \@firstoftwo
 \else \expandafter \@secondoftwo
 \fi
}%
\providecommand \natexlab [1]{#1}%
\providecommand \enquote  [1]{``#1''}%
\providecommand \bibnamefont  [1]{#1}%
\providecommand \bibfnamefont [1]{#1}%
\providecommand \citenamefont [1]{#1}%
\providecommand \href@noop [0]{\@secondoftwo}%
\providecommand \href [0]{\begingroup \@sanitize@url \@href}%
\providecommand \@href[1]{\@@startlink{#1}\@@href}%
\providecommand \@@href[1]{\endgroup#1\@@endlink}%
\providecommand \@sanitize@url [0]{\catcode `\\12\catcode `\$12\catcode
  `\&12\catcode `\#12\catcode `\^12\catcode `\_12\catcode `\%12\relax}%
\providecommand \@@startlink[1]{}%
\providecommand \@@endlink[0]{}%
\providecommand \url  [0]{\begingroup\@sanitize@url \@url }%
\providecommand \@url [1]{\endgroup\@href {#1}{\urlprefix }}%
\providecommand \urlprefix  [0]{URL }%
\providecommand \Eprint [0]{\href }%
\providecommand \doibase [0]{http://dx.doi.org/}%
\providecommand \selectlanguage [0]{\@gobble}%
\providecommand \bibinfo  [0]{\@secondoftwo}%
\providecommand \bibfield  [0]{\@secondoftwo}%
\providecommand \translation [1]{[#1]}%
\providecommand \BibitemOpen [0]{}%
\providecommand \bibitemStop [0]{}%
\providecommand \bibitemNoStop [0]{.\EOS\space}%
\providecommand \EOS [0]{\spacefactor3000\relax}%
\providecommand \BibitemShut  [1]{\csname bibitem#1\endcsname}%
\let\auto@bib@innerbib\@empty
\bibitem [{\citenamefont {Lopes dos Santos}\ \emph
  {et~al.}(2007)\citenamefont {Lopes dos Santos}, \citenamefont {Peres},\
  and\ \citenamefont {Castro Neto}}]{lopesdossantos_graphene_2007}%
  \BibitemOpen
  \bibfield  {author} {\bibinfo {author} {\bibfnamefont {J.~M.~B.}\
  \bibnamefont {Lopes dos Santos}}, \bibinfo {author} {\bibfnamefont
  {N.~M.~R.}\ \bibnamefont {Peres}}, \ and\ \bibinfo {author} {\bibfnamefont
  {A.~H.}\ \bibnamefont {Castro Neto}},\ }\href {\doibase
  10.1103/PhysRevLett.99.256802} {\bibfield  {journal} {\bibinfo  {journal}
  {Phys. Rev. Lett.}\ }\textbf {\bibinfo {volume} {99}},\ \bibinfo {pages}
  {256802} (\bibinfo {year} {2007})}\BibitemShut {NoStop}%
\bibitem [{\citenamefont {Trambly~de Laissardière}\ \emph
  {et~al.}(2010)\citenamefont {Trambly~de Laissardière}, \citenamefont
  {Mayou},\ and\ \citenamefont
  {Magaud}}]{trambly_de_laissardiere_localization_2010}%
  \BibitemOpen
  \bibfield  {author} {\bibinfo {author} {\bibfnamefont {G.}~\bibnamefont
  {Trambly~de Laissardière}}, \bibinfo {author} {\bibfnamefont
  {D.}~\bibnamefont {Mayou}}, \ and\ \bibinfo {author} {\bibfnamefont
  {L.}~\bibnamefont {Magaud}},\ }\href {\doibase 10.1021/nl902948m} {\bibfield
  {journal} {\bibinfo  {journal} {Nano Lett.}\ }\textbf {\bibinfo {volume}
  {10}},\ \bibinfo {pages} {804} (\bibinfo {year} {2010})}\BibitemShut
  {NoStop}%
\bibitem [{\citenamefont {Bistritzer}\ and\ \citenamefont
  {MacDonald}(2011{\natexlab{a}})}]{bistritzer_moire_2011}%
  \BibitemOpen
  \bibfield  {author} {\bibinfo {author} {\bibfnamefont {R.}~\bibnamefont
  {Bistritzer}}\ and\ \bibinfo {author} {\bibfnamefont {A.~H.}\ \bibnamefont
  {MacDonald}},\ }\href {\doibase 10.1073/pnas.1108174108} {\bibfield
  {journal} {\bibinfo  {journal} {Proceedings of the National Academy of
  Sciences}\ }\textbf {\bibinfo {volume} {108}},\ \bibinfo {pages} {12233}
  (\bibinfo {year} {2011}{\natexlab{a}})}\BibitemShut {NoStop}%
\bibitem [{\citenamefont {Carr}\ \emph {et~al.}(2020)\citenamefont {Carr},
  \citenamefont {Fang},\ and\ \citenamefont
  {Kaxiras}}]{carr_electronic-structure_2020}%
  \BibitemOpen
  \bibfield  {author} {\bibinfo {author} {\bibfnamefont {S.}~\bibnamefont
  {Carr}}, \bibinfo {author} {\bibfnamefont {S.}~\bibnamefont {Fang}}, \ and\
  \bibinfo {author} {\bibfnamefont {E.}~\bibnamefont {Kaxiras}},\ }\href
  {\doibase 10.1038/s41578-020-0214-0} {\bibfield  {journal} {\bibinfo
  {journal} {Nature Reviews Materials}\ }\textbf {\bibinfo {volume} {5}},\
  \bibinfo {pages} {748} (\bibinfo {year} {2020})}\BibitemShut {NoStop}%
\bibitem [{\citenamefont {Cao}\ \emph {et~al.}(2018{\natexlab{a}})\citenamefont
  {Cao}, \citenamefont {Fatemi}, \citenamefont {Fang}, \citenamefont
  {Watanabe}, \citenamefont {Taniguchi}, \citenamefont {Kaxiras},\ and\
  \citenamefont {Jarillo-Herrero}}]{cao_unconventional_2018}%
  \BibitemOpen
  \bibfield  {author} {\bibinfo {author} {\bibfnamefont {Y.}~\bibnamefont
  {Cao}}, \bibinfo {author} {\bibfnamefont {V.}~\bibnamefont {Fatemi}},
  \bibinfo {author} {\bibfnamefont {S.}~\bibnamefont {Fang}}, \bibinfo {author}
  {\bibfnamefont {K.}~\bibnamefont {Watanabe}}, \bibinfo {author}
  {\bibfnamefont {T.}~\bibnamefont {Taniguchi}}, \bibinfo {author}
  {\bibfnamefont {E.}~\bibnamefont {Kaxiras}}, \ and\ \bibinfo {author}
  {\bibfnamefont {P.}~\bibnamefont {Jarillo-Herrero}},\ }\href {\doibase
  10.1038/nature26160} {\bibfield  {journal} {\bibinfo  {journal} {Nature}\
  }\textbf {\bibinfo {volume} {556}},\ \bibinfo {pages} {43} (\bibinfo {year}
  {2018}{\natexlab{a}})}\BibitemShut {NoStop}%
\bibitem [{\citenamefont {Cao}\ \emph {et~al.}(2018{\natexlab{b}})\citenamefont
  {Cao}, \citenamefont {Fatemi}, \citenamefont {Demir}, \citenamefont {Fang},
  \citenamefont {Tomarken}, \citenamefont {Luo}, \citenamefont
  {Sanchez-Yamagishi}, \citenamefont {Watanabe}, \citenamefont {Taniguchi},
  \citenamefont {Kaxiras}, \citenamefont {Ashoori},\ and\ \citenamefont
  {Jarillo-Herrero}}]{cao_correlated_2018}%
  \BibitemOpen
  \bibfield  {author} {\bibinfo {author} {\bibfnamefont {Y.}~\bibnamefont
  {Cao}}, \bibinfo {author} {\bibfnamefont {V.}~\bibnamefont {Fatemi}},
  \bibinfo {author} {\bibfnamefont {A.}~\bibnamefont {Demir}}, \bibinfo
  {author} {\bibfnamefont {S.}~\bibnamefont {Fang}}, \bibinfo {author}
  {\bibfnamefont {S.~L.}\ \bibnamefont {Tomarken}}, \bibinfo {author}
  {\bibfnamefont {J.~Y.}\ \bibnamefont {Luo}}, \bibinfo {author} {\bibfnamefont
  {J.~D.}\ \bibnamefont {Sanchez-Yamagishi}}, \bibinfo {author} {\bibfnamefont
  {K.}~\bibnamefont {Watanabe}}, \bibinfo {author} {\bibfnamefont
  {T.}~\bibnamefont {Taniguchi}}, \bibinfo {author} {\bibfnamefont
  {E.}~\bibnamefont {Kaxiras}}, \bibinfo {author} {\bibfnamefont {R.~C.}\
  \bibnamefont {Ashoori}}, \ and\ \bibinfo {author} {\bibfnamefont
  {P.}~\bibnamefont {Jarillo-Herrero}},\ }\href {\doibase 10.1038/nature26154}
  {\bibfield  {journal} {\bibinfo  {journal} {Nature}\ }\textbf {\bibinfo
  {volume} {556}},\ \bibinfo {pages} {80} (\bibinfo {year}
  {2018}{\natexlab{b}})}\BibitemShut {NoStop}%
\bibitem [{\citenamefont {Chen}\ \emph {et~al.}(2019)\citenamefont {Chen},
  \citenamefont {Jiang}, \citenamefont {Wu}, \citenamefont {Lyu}, \citenamefont
  {Li}, \citenamefont {Chittari}, \citenamefont {Watanabe}, \citenamefont
  {Taniguchi}, \citenamefont {Shi}, \citenamefont {Jung}, \citenamefont
  {Zhang},\ and\ \citenamefont {Wang}}]{chen_evidence_2019}%
  \BibitemOpen
  \bibfield  {author} {\bibinfo {author} {\bibfnamefont {G.}~\bibnamefont
  {Chen}}, \bibinfo {author} {\bibfnamefont {L.}~\bibnamefont {Jiang}},
  \bibinfo {author} {\bibfnamefont {S.}~\bibnamefont {Wu}}, \bibinfo {author}
  {\bibfnamefont {B.}~\bibnamefont {Lyu}}, \bibinfo {author} {\bibfnamefont
  {H.}~\bibnamefont {Li}}, \bibinfo {author} {\bibfnamefont {B.~L.}\
  \bibnamefont {Chittari}}, \bibinfo {author} {\bibfnamefont {K.}~\bibnamefont
  {Watanabe}}, \bibinfo {author} {\bibfnamefont {T.}~\bibnamefont {Taniguchi}},
  \bibinfo {author} {\bibfnamefont {Z.}~\bibnamefont {Shi}}, \bibinfo {author}
  {\bibfnamefont {J.}~\bibnamefont {Jung}}, \bibinfo {author} {\bibfnamefont
  {Y.}~\bibnamefont {Zhang}}, \ and\ \bibinfo {author} {\bibfnamefont
  {F.}~\bibnamefont {Wang}},\ }\href {\doibase 10.1038/s41567-018-0387-2}
  {\bibfield  {journal} {\bibinfo  {journal} {Nat. Phys.}\ }\textbf {\bibinfo
  {volume} {15}},\ \bibinfo {pages} {237} (\bibinfo {year} {2019})}\BibitemShut
  {NoStop}%
\bibitem [{\citenamefont {Xie}\ \emph {et~al.}(2021)\citenamefont {Xie},
  \citenamefont {Pierce}, \citenamefont {Park}, \citenamefont {Parker},
  \citenamefont {Khalaf}, \citenamefont {Ledwith}, \citenamefont {Cao},
  \citenamefont {Lee}, \citenamefont {Chen}, \citenamefont {Forrester},
  \citenamefont {Watanabe}, \citenamefont {Taniguchi}, \citenamefont
  {Vishwanath}, \citenamefont {Jarillo-Herrero},\ and\ \citenamefont
  {Yacoby}}]{xie_fractional_2021}%
  \BibitemOpen
  \bibfield  {author} {\bibinfo {author} {\bibfnamefont {Y.}~\bibnamefont
  {Xie}}, \bibinfo {author} {\bibfnamefont {A.~T.}\ \bibnamefont {Pierce}},
  \bibinfo {author} {\bibfnamefont {J.~M.}\ \bibnamefont {Park}}, \bibinfo
  {author} {\bibfnamefont {D.~E.}\ \bibnamefont {Parker}}, \bibinfo {author}
  {\bibfnamefont {E.}~\bibnamefont {Khalaf}}, \bibinfo {author} {\bibfnamefont
  {P.}~\bibnamefont {Ledwith}}, \bibinfo {author} {\bibfnamefont
  {Y.}~\bibnamefont {Cao}}, \bibinfo {author} {\bibfnamefont {S.~H.}\
  \bibnamefont {Lee}}, \bibinfo {author} {\bibfnamefont {S.}~\bibnamefont
  {Chen}}, \bibinfo {author} {\bibfnamefont {P.~R.}\ \bibnamefont {Forrester}},
  \bibinfo {author} {\bibfnamefont {K.}~\bibnamefont {Watanabe}}, \bibinfo
  {author} {\bibfnamefont {T.}~\bibnamefont {Taniguchi}}, \bibinfo {author}
  {\bibfnamefont {A.}~\bibnamefont {Vishwanath}}, \bibinfo {author}
  {\bibfnamefont {P.}~\bibnamefont {Jarillo-Herrero}}, \ and\ \bibinfo {author}
  {\bibfnamefont {A.}~\bibnamefont {Yacoby}},\ }\href {\doibase
  10.1038/s41586-021-04002-3} {\bibfield  {journal} {\bibinfo  {journal}
  {Nature}\ }\textbf {\bibinfo {volume} {600}},\ \bibinfo {pages} {439}
  (\bibinfo {year} {2021})}\BibitemShut {NoStop}%
\bibitem [{\citenamefont {Bistritzer}\ and\ \citenamefont
  {MacDonald}(2011{\natexlab{b}})}]{bistritzer_butterfly_2011}%
  \BibitemOpen
  \bibfield  {author} {\bibinfo {author} {\bibfnamefont {R.}~\bibnamefont
  {Bistritzer}}\ and\ \bibinfo {author} {\bibfnamefont {A.~H.}\ \bibnamefont
  {MacDonald}},\ }\href {\doibase 10.1103/PhysRevB.84.035440} {\bibfield
  {journal} {\bibinfo  {journal} {Phys. Rev. B}\ }\textbf {\bibinfo {volume}
  {84}},\ \bibinfo {pages} {035440} (\bibinfo {year}
  {2011}{\natexlab{b}})}\BibitemShut {NoStop}%
\bibitem [{\citenamefont {Dean}\ \emph {et~al.}(2013)\citenamefont {Dean},
  \citenamefont {Wang}, \citenamefont {Maher}, \citenamefont {Forsythe},
  \citenamefont {Ghahari}, \citenamefont {Gao}, \citenamefont {Katoch},
  \citenamefont {Ishigami}, \citenamefont {Moon}, \citenamefont {Koshino},
  \citenamefont {Taniguchi}, \citenamefont {Watanabe}, \citenamefont {Shepard},
  \citenamefont {Hone},\ and\ \citenamefont {Kim}}]{dean_hofstadters_2013}%
  \BibitemOpen
  \bibfield  {author} {\bibinfo {author} {\bibfnamefont {C.~R.}\ \bibnamefont
  {Dean}}, \bibinfo {author} {\bibfnamefont {L.}~\bibnamefont {Wang}}, \bibinfo
  {author} {\bibfnamefont {P.}~\bibnamefont {Maher}}, \bibinfo {author}
  {\bibfnamefont {C.}~\bibnamefont {Forsythe}}, \bibinfo {author}
  {\bibfnamefont {F.}~\bibnamefont {Ghahari}}, \bibinfo {author} {\bibfnamefont
  {Y.}~\bibnamefont {Gao}}, \bibinfo {author} {\bibfnamefont {J.}~\bibnamefont
  {Katoch}}, \bibinfo {author} {\bibfnamefont {M.}~\bibnamefont {Ishigami}},
  \bibinfo {author} {\bibfnamefont {P.}~\bibnamefont {Moon}}, \bibinfo {author}
  {\bibfnamefont {M.}~\bibnamefont {Koshino}}, \bibinfo {author} {\bibfnamefont
  {T.}~\bibnamefont {Taniguchi}}, \bibinfo {author} {\bibfnamefont
  {K.}~\bibnamefont {Watanabe}}, \bibinfo {author} {\bibfnamefont {K.~L.}\
  \bibnamefont {Shepard}}, \bibinfo {author} {\bibfnamefont {J.}~\bibnamefont
  {Hone}}, \ and\ \bibinfo {author} {\bibfnamefont {P.}~\bibnamefont {Kim}},\
  }\href {\doibase 10.1038/nature12186} {\bibfield  {journal} {\bibinfo
  {journal} {Nature}\ }\textbf {\bibinfo {volume} {497}},\ \bibinfo {pages}
  {598} (\bibinfo {year} {2013})}\BibitemShut {NoStop}%
\bibitem [{\citenamefont {Hunt}\ \emph {et~al.}(2013)\citenamefont {Hunt},
  \citenamefont {Sanchez-Yamagishi}, \citenamefont {Young}, \citenamefont
  {Yankowitz}, \citenamefont {LeRoy}, \citenamefont {Watanabe}, \citenamefont
  {Taniguchi}, \citenamefont {Moon}, \citenamefont {Koshino}, \citenamefont
  {Jarillo-Herrero},\ and\ \citenamefont {Ashoori}}]{hunt_massive_2013}%
  \BibitemOpen
  \bibfield  {author} {\bibinfo {author} {\bibfnamefont {B.}~\bibnamefont
  {Hunt}}, \bibinfo {author} {\bibfnamefont {J.~D.}\ \bibnamefont
  {Sanchez-Yamagishi}}, \bibinfo {author} {\bibfnamefont {A.~F.}\ \bibnamefont
  {Young}}, \bibinfo {author} {\bibfnamefont {M.}~\bibnamefont {Yankowitz}},
  \bibinfo {author} {\bibfnamefont {B.~J.}\ \bibnamefont {LeRoy}}, \bibinfo
  {author} {\bibfnamefont {K.}~\bibnamefont {Watanabe}}, \bibinfo {author}
  {\bibfnamefont {T.}~\bibnamefont {Taniguchi}}, \bibinfo {author}
  {\bibfnamefont {P.}~\bibnamefont {Moon}}, \bibinfo {author} {\bibfnamefont
  {M.}~\bibnamefont {Koshino}}, \bibinfo {author} {\bibfnamefont
  {P.}~\bibnamefont {Jarillo-Herrero}}, \ and\ \bibinfo {author} {\bibfnamefont
  {R.~C.}\ \bibnamefont {Ashoori}},\ }\href {\doibase 10.1126/science.1237240}
  {\bibfield  {journal} {\bibinfo  {journal} {Science}\ }\textbf {\bibinfo
  {volume} {340}},\ \bibinfo {pages} {1427} (\bibinfo {year}
  {2013})}\BibitemShut {NoStop}%
\bibitem [{\citenamefont {Ponomarenko}\ \emph {et~al.}(2013)\citenamefont
  {Ponomarenko}, \citenamefont {Gorbachev}, \citenamefont {Yu}, \citenamefont
  {Elias}, \citenamefont {Jalil}, \citenamefont {Patel}, \citenamefont
  {Mishchenko}, \citenamefont {Mayorov}, \citenamefont {Woods}, \citenamefont
  {Wallbank}, \citenamefont {Mucha-Kruczynski}, \citenamefont {Piot},
  \citenamefont {Potemski}, \citenamefont {Grigorieva}, \citenamefont
  {Novoselov}, \citenamefont {Guinea}, \citenamefont {Fal’ko},\ and\
  \citenamefont {Geim}}]{ponomarenko_cloning_2013}%
  \BibitemOpen
  \bibfield  {author} {\bibinfo {author} {\bibfnamefont {L.~A.}\ \bibnamefont
  {Ponomarenko}}, \bibinfo {author} {\bibfnamefont {R.~V.}\ \bibnamefont
  {Gorbachev}}, \bibinfo {author} {\bibfnamefont {G.~L.}\ \bibnamefont {Yu}},
  \bibinfo {author} {\bibfnamefont {D.~C.}\ \bibnamefont {Elias}}, \bibinfo
  {author} {\bibfnamefont {R.}~\bibnamefont {Jalil}}, \bibinfo {author}
  {\bibfnamefont {A.~A.}\ \bibnamefont {Patel}}, \bibinfo {author}
  {\bibfnamefont {A.}~\bibnamefont {Mishchenko}}, \bibinfo {author}
  {\bibfnamefont {A.~S.}\ \bibnamefont {Mayorov}}, \bibinfo {author}
  {\bibfnamefont {C.~R.}\ \bibnamefont {Woods}}, \bibinfo {author}
  {\bibfnamefont {J.~R.}\ \bibnamefont {Wallbank}}, \bibinfo {author}
  {\bibfnamefont {M.}~\bibnamefont {Mucha-Kruczynski}}, \bibinfo {author}
  {\bibfnamefont {B.~A.}\ \bibnamefont {Piot}}, \bibinfo {author}
  {\bibfnamefont {M.}~\bibnamefont {Potemski}}, \bibinfo {author}
  {\bibfnamefont {I.~V.}\ \bibnamefont {Grigorieva}}, \bibinfo {author}
  {\bibfnamefont {K.~S.}\ \bibnamefont {Novoselov}}, \bibinfo {author}
  {\bibfnamefont {F.}~\bibnamefont {Guinea}}, \bibinfo {author} {\bibfnamefont
  {V.~I.}\ \bibnamefont {Fal’ko}}, \ and\ \bibinfo {author} {\bibfnamefont
  {A.~K.}\ \bibnamefont {Geim}},\ }\href {\doibase 10.1038/nature12187}
  {\bibfield  {journal} {\bibinfo  {journal} {Nature}\ }\textbf {\bibinfo
  {volume} {497}},\ \bibinfo {pages} {594} (\bibinfo {year}
  {2013})}\BibitemShut {NoStop}%
\bibitem [{\citenamefont {Zhang}\ and\ \citenamefont
  {Shi}(2014)}]{zhang_quantum_2014}%
  \BibitemOpen
  \bibfield  {author} {\bibinfo {author} {\bibfnamefont {Y.}~\bibnamefont
  {Zhang}}\ and\ \bibinfo {author} {\bibfnamefont {J.}~\bibnamefont {Shi}},\
  }\href {\doibase 10.1103/PhysRevLett.113.016801} {\bibfield  {journal}
  {\bibinfo  {journal} {Phys. Rev. Lett.}\ }\textbf {\bibinfo {volume} {113}},\
  \bibinfo {pages} {016801} (\bibinfo {year} {2014})}\BibitemShut {NoStop}%
\bibitem [{\citenamefont {Jain}(1989)}]{jain_composite-fermion_1989}%
  \BibitemOpen
  \bibfield  {author} {\bibinfo {author} {\bibfnamefont {J.~K.}\ \bibnamefont
  {Jain}},\ }\href {\doibase 10.1103/PhysRevLett.63.199} {\bibfield  {journal}
  {\bibinfo  {journal} {Phys. Rev. Lett.}\ }\textbf {\bibinfo {volume} {63}},\
  \bibinfo {pages} {199} (\bibinfo {year} {1989})}\BibitemShut {NoStop}%
\bibitem [{\citenamefont {Halperin}\ \emph {et~al.}(1993)\citenamefont
  {Halperin}, \citenamefont {Lee},\ and\ \citenamefont
  {Read}}]{halperin_theory_1993}%
  \BibitemOpen
  \bibfield  {author} {\bibinfo {author} {\bibfnamefont {B.~I.}\ \bibnamefont
  {Halperin}}, \bibinfo {author} {\bibfnamefont {P.~A.}\ \bibnamefont {Lee}}, \
  and\ \bibinfo {author} {\bibfnamefont {N.}~\bibnamefont {Read}},\ }\href
  {\doibase 10.1103/PhysRevB.47.7312} {\bibfield  {journal} {\bibinfo
  {journal} {Phys. Rev. B}\ }\textbf {\bibinfo {volume} {47}},\ \bibinfo
  {pages} {7312} (\bibinfo {year} {1993})}\BibitemShut {NoStop}%
\bibitem [{\citenamefont {Li}\ \emph {et~al.}(2017)\citenamefont {Li},
  \citenamefont {Tan}, \citenamefont {Chen}, \citenamefont {Zeng},
  \citenamefont {Taniguchi}, \citenamefont {Watanabe}, \citenamefont {Hone},\
  and\ \citenamefont {Dean}}]{li_even-denominator_2017}%
  \BibitemOpen
  \bibfield  {author} {\bibinfo {author} {\bibfnamefont {J.~I.~A.}\
  \bibnamefont {Li}}, \bibinfo {author} {\bibfnamefont {C.}~\bibnamefont
  {Tan}}, \bibinfo {author} {\bibfnamefont {S.}~\bibnamefont {Chen}}, \bibinfo
  {author} {\bibfnamefont {Y.}~\bibnamefont {Zeng}}, \bibinfo {author}
  {\bibfnamefont {T.}~\bibnamefont {Taniguchi}}, \bibinfo {author}
  {\bibfnamefont {K.}~\bibnamefont {Watanabe}}, \bibinfo {author}
  {\bibfnamefont {J.}~\bibnamefont {Hone}}, \ and\ \bibinfo {author}
  {\bibfnamefont {C.~R.}\ \bibnamefont {Dean}},\ }\href {\doibase
  10.1126/science.aao2521} {\bibfield  {journal} {\bibinfo  {journal}
  {Science}\ }\textbf {\bibinfo {volume} {358}},\ \bibinfo {pages} {648}
  (\bibinfo {year} {2017})}\BibitemShut {NoStop}%
\bibitem [{\citenamefont {Zibrov}\ \emph {et~al.}(2017)\citenamefont {Zibrov},
  \citenamefont {Kometter}, \citenamefont {Zhou}, \citenamefont {Spanton},
  \citenamefont {Taniguchi}, \citenamefont {Watanabe}, \citenamefont
  {Zaletel},\ and\ \citenamefont {Young}}]{zibrov_tunable_2017}%
  \BibitemOpen
  \bibfield  {author} {\bibinfo {author} {\bibfnamefont {A.~A.}\ \bibnamefont
  {Zibrov}}, \bibinfo {author} {\bibfnamefont {C.}~\bibnamefont {Kometter}},
  \bibinfo {author} {\bibfnamefont {H.}~\bibnamefont {Zhou}}, \bibinfo {author}
  {\bibfnamefont {E.~M.}\ \bibnamefont {Spanton}}, \bibinfo {author}
  {\bibfnamefont {T.}~\bibnamefont {Taniguchi}}, \bibinfo {author}
  {\bibfnamefont {K.}~\bibnamefont {Watanabe}}, \bibinfo {author}
  {\bibfnamefont {M.~P.}\ \bibnamefont {Zaletel}}, \ and\ \bibinfo {author}
  {\bibfnamefont {A.~F.}\ \bibnamefont {Young}},\ }\href {\doibase
  10.1038/nature23893} {\bibfield  {journal} {\bibinfo  {journal} {Nature}\
  }\textbf {\bibinfo {volume} {549}},\ \bibinfo {pages} {360} (\bibinfo {year}
  {2017})}\BibitemShut {NoStop}%
\bibitem [{\citenamefont {Haldane}(1988)}]{haldane_model_1988}%
  \BibitemOpen
  \bibfield  {author} {\bibinfo {author} {\bibfnamefont {F.~D.~M.}\
  \bibnamefont {Haldane}},\ }\href {\doibase 10.1103/PhysRevLett.61.2015}
  {\bibfield  {journal} {\bibinfo  {journal} {Phys. Rev. Lett.}\ }\textbf
  {\bibinfo {volume} {61}},\ \bibinfo {pages} {2015} (\bibinfo {year}
  {1988})}\BibitemShut {NoStop}%
\bibitem [{\citenamefont {Jain}(2007)}]{jain2007composite}%
  \BibitemOpen
  \bibfield  {author} {\bibinfo {author} {\bibfnamefont {J.}~\bibnamefont
  {Jain}},\ }\href {https://books.google.com.hk/books?id=0jv9UF6UL20C} {\emph
  {\bibinfo {title} {Composite Fermions}}}\ (\bibinfo  {publisher} {Cambridge
  University Press},\ \bibinfo {year} {2007})\BibitemShut {NoStop}%
\bibitem [{\citenamefont {Seiberg}\ \emph {et~al.}(2016)\citenamefont
  {Seiberg}, \citenamefont {Senthil}, \citenamefont {Wang},\ and\ \citenamefont
  {Witten}}]{seiberg_duality_2016}%
  \BibitemOpen
  \bibfield  {author} {\bibinfo {author} {\bibfnamefont {N.}~\bibnamefont
  {Seiberg}}, \bibinfo {author} {\bibfnamefont {T.}~\bibnamefont {Senthil}},
  \bibinfo {author} {\bibfnamefont {C.}~\bibnamefont {Wang}}, \ and\ \bibinfo
  {author} {\bibfnamefont {E.}~\bibnamefont {Witten}},\ }\href {\doibase
  10.1016/j.aop.2016.08.007} {\bibfield  {journal} {\bibinfo  {journal} {Annals
  of Physics}\ }\textbf {\bibinfo {volume} {374}},\ \bibinfo {pages} {395}
  (\bibinfo {year} {2016})}\BibitemShut {NoStop}%
\bibitem [{\citenamefont {Senthil}\ \emph {et~al.}(2019)\citenamefont
  {Senthil}, \citenamefont {Son}, \citenamefont {Wang},\ and\ \citenamefont
  {Xu}}]{senthil_duality_2019}%
  \BibitemOpen
  \bibfield  {author} {\bibinfo {author} {\bibfnamefont {T.}~\bibnamefont
  {Senthil}}, \bibinfo {author} {\bibfnamefont {D.~T.}\ \bibnamefont {Son}},
  \bibinfo {author} {\bibfnamefont {C.}~\bibnamefont {Wang}}, \ and\ \bibinfo
  {author} {\bibfnamefont {C.}~\bibnamefont {Xu}},\ }\href {\doibase
  https://doi.org/10.1016/j.physrep.2019.09.001} {\bibfield  {journal}
  {\bibinfo  {journal} {Physics Reports}\ }\textbf {\bibinfo {volume} {827}},\
  \bibinfo {pages} {1} (\bibinfo {year} {2019})}\BibitemShut {NoStop}%
\bibitem [{\citenamefont {Son}(2015)}]{son_is_2015}%
  \BibitemOpen
  \bibfield  {author} {\bibinfo {author} {\bibfnamefont {D.~T.}\ \bibnamefont
  {Son}},\ }\href {\doibase 10.1103/PhysRevX.5.031027} {\bibfield  {journal}
  {\bibinfo  {journal} {Phys. Rev. X}\ }\textbf {\bibinfo {volume} {5}},\
  \bibinfo {pages} {031027} (\bibinfo {year} {2015})}\BibitemShut {NoStop}%
\bibitem [{\citenamefont {Zhang}\ \emph {et~al.}(2009)\citenamefont {Zhang},
  \citenamefont {Tang}, \citenamefont {Girit}, \citenamefont {Hao},
  \citenamefont {Martin}, \citenamefont {Zettl}, \citenamefont {Crommie},
  \citenamefont {Shen},\ and\ \citenamefont {Wang}}]{zhang_direct_2009}%
  \BibitemOpen
  \bibfield  {author} {\bibinfo {author} {\bibfnamefont {Y.}~\bibnamefont
  {Zhang}}, \bibinfo {author} {\bibfnamefont {T.-T.}\ \bibnamefont {Tang}},
  \bibinfo {author} {\bibfnamefont {C.}~\bibnamefont {Girit}}, \bibinfo
  {author} {\bibfnamefont {Z.}~\bibnamefont {Hao}}, \bibinfo {author}
  {\bibfnamefont {M.~C.}\ \bibnamefont {Martin}}, \bibinfo {author}
  {\bibfnamefont {A.}~\bibnamefont {Zettl}}, \bibinfo {author} {\bibfnamefont
  {M.~F.}\ \bibnamefont {Crommie}}, \bibinfo {author} {\bibfnamefont {Y.~R.}\
  \bibnamefont {Shen}}, \ and\ \bibinfo {author} {\bibfnamefont
  {F.}~\bibnamefont {Wang}},\ }\href {\doibase 10.1038/nature08105} {\bibfield
  {journal} {\bibinfo  {journal} {Nature}\ }\textbf {\bibinfo {volume} {459}},\
  \bibinfo {pages} {820} (\bibinfo {year} {2009})}\BibitemShut {NoStop}%
\bibitem [{\citenamefont {Spanton}\ \emph {et~al.}(2018)\citenamefont
  {Spanton}, \citenamefont {Zibrov}, \citenamefont {Zhou}, \citenamefont
  {Taniguchi}, \citenamefont {Watanabe}, \citenamefont {Zaletel},\ and\
  \citenamefont {Young}}]{spanton_observation_2018}%
  \BibitemOpen
  \bibfield  {author} {\bibinfo {author} {\bibfnamefont {E.~M.}\ \bibnamefont
  {Spanton}}, \bibinfo {author} {\bibfnamefont {A.~A.}\ \bibnamefont {Zibrov}},
  \bibinfo {author} {\bibfnamefont {H.}~\bibnamefont {Zhou}}, \bibinfo {author}
  {\bibfnamefont {T.}~\bibnamefont {Taniguchi}}, \bibinfo {author}
  {\bibfnamefont {K.}~\bibnamefont {Watanabe}}, \bibinfo {author}
  {\bibfnamefont {M.~P.}\ \bibnamefont {Zaletel}}, \ and\ \bibinfo {author}
  {\bibfnamefont {A.~F.}\ \bibnamefont {Young}},\ }\href {\doibase
  10.1126/science.aan8458} {\bibfield  {journal} {\bibinfo  {journal}
  {Science}\ }\textbf {\bibinfo {volume} {360}},\ \bibinfo {pages} {62}
  (\bibinfo {year} {2018})}\BibitemShut {NoStop}%
\bibitem [{\citenamefont {Polshyn}\ \emph {et~al.}(2022)\citenamefont
  {Polshyn}, \citenamefont {Zhang}, \citenamefont {Kumar}, \citenamefont
  {Soejima}, \citenamefont {Ledwith}, \citenamefont {Watanabe}, \citenamefont
  {Taniguchi}, \citenamefont {Vishwanath}, \citenamefont {Zaletel},\ and\
  \citenamefont {Young}}]{polshyn_topological_2021}%
  \BibitemOpen
  \bibfield  {author} {\bibinfo {author} {\bibfnamefont {H.}~\bibnamefont
  {Polshyn}}, \bibinfo {author} {\bibfnamefont {Y.}~\bibnamefont {Zhang}},
  \bibinfo {author} {\bibfnamefont {M.~A.}\ \bibnamefont {Kumar}}, \bibinfo
  {author} {\bibfnamefont {T.}~\bibnamefont {Soejima}}, \bibinfo {author}
  {\bibfnamefont {P.}~\bibnamefont {Ledwith}}, \bibinfo {author} {\bibfnamefont
  {K.}~\bibnamefont {Watanabe}}, \bibinfo {author} {\bibfnamefont
  {T.}~\bibnamefont {Taniguchi}}, \bibinfo {author} {\bibfnamefont
  {A.}~\bibnamefont {Vishwanath}}, \bibinfo {author} {\bibfnamefont {M.~P.}\
  \bibnamefont {Zaletel}}, \ and\ \bibinfo {author} {\bibfnamefont {A.~F.}\
  \bibnamefont {Young}},\ }\href@noop {} {\bibfield  {journal} {\bibinfo
  {journal} {Nature Physics}\ }\textbf {\bibinfo {volume} {18}},\ \bibinfo
  {pages} {42} (\bibinfo {year} {2022})}\BibitemShut {NoStop}%
\bibitem [{\citenamefont {Moon}\ and\ \citenamefont
  {Koshino}(2012)}]{moon_energy_2012}%
  \BibitemOpen
  \bibfield  {author} {\bibinfo {author} {\bibfnamefont {P.}~\bibnamefont
  {Moon}}\ and\ \bibinfo {author} {\bibfnamefont {M.}~\bibnamefont {Koshino}},\
  }\href {\doibase 10.1103/PhysRevB.85.195458} {\bibfield  {journal} {\bibinfo
  {journal} {Phys. Rev. B}\ }\textbf {\bibinfo {volume} {85}},\ \bibinfo
  {pages} {195458} (\bibinfo {year} {2012})}\BibitemShut {NoStop}%
\bibitem [{\citenamefont {Castro~Neto}\ \emph {et~al.}(2009)\citenamefont
  {Castro~Neto}, \citenamefont {Guinea}, \citenamefont {Peres}, \citenamefont
  {Novoselov},\ and\ \citenamefont {Geim}}]{castro_neto_electronic_2009}%
  \BibitemOpen
  \bibfield  {author} {\bibinfo {author} {\bibfnamefont {A.~H.}\ \bibnamefont
  {Castro~Neto}}, \bibinfo {author} {\bibfnamefont {F.}~\bibnamefont {Guinea}},
  \bibinfo {author} {\bibfnamefont {N.~M.~R.}\ \bibnamefont {Peres}}, \bibinfo
  {author} {\bibfnamefont {K.~S.}\ \bibnamefont {Novoselov}}, \ and\ \bibinfo
  {author} {\bibfnamefont {A.~K.}\ \bibnamefont {Geim}},\ }\href {\doibase
  10.1103/RevModPhys.81.109} {\bibfield  {journal} {\bibinfo  {journal} {Rev.
  Mod. Phys.}\ }\textbf {\bibinfo {volume} {81}},\ \bibinfo {pages} {109}
  (\bibinfo {year} {2009})}\BibitemShut {NoStop}%
\bibitem [{\citenamefont {Huang}\ \emph {et~al.}(2022)\citenamefont {Huang},
  \citenamefont {Fu}, \citenamefont {Hickey}, \citenamefont {Alem},
  \citenamefont {Lin}, \citenamefont {Watanabe}, \citenamefont {Taniguchi},\
  and\ \citenamefont {Zhu}}]{huang_valley_2022}%
  \BibitemOpen
  \bibfield  {author} {\bibinfo {author} {\bibfnamefont {K.}~\bibnamefont
  {Huang}}, \bibinfo {author} {\bibfnamefont {H.}~\bibnamefont {Fu}}, \bibinfo
  {author} {\bibfnamefont {D.~R.}\ \bibnamefont {Hickey}}, \bibinfo {author}
  {\bibfnamefont {N.}~\bibnamefont {Alem}}, \bibinfo {author} {\bibfnamefont
  {X.}~\bibnamefont {Lin}}, \bibinfo {author} {\bibfnamefont {K.}~\bibnamefont
  {Watanabe}}, \bibinfo {author} {\bibfnamefont {T.}~\bibnamefont {Taniguchi}},
  \ and\ \bibinfo {author} {\bibfnamefont {J.}~\bibnamefont {Zhu}},\ }\href
  {\doibase 10.1103/PhysRevX.12.031019} {\bibfield  {journal} {\bibinfo
  {journal} {Phys. Rev. X}\ }\textbf {\bibinfo {volume} {12}},\ \bibinfo
  {pages} {031019} (\bibinfo {year} {2022})}\BibitemShut {NoStop}%
\bibitem [{\citenamefont {Girvin}\ and\ \citenamefont
  {Jach}(1984)}]{girvin_formalism_1984}%
  \BibitemOpen
  \bibfield  {author} {\bibinfo {author} {\bibfnamefont {S.~M.}\ \bibnamefont
  {Girvin}}\ and\ \bibinfo {author} {\bibfnamefont {T.}~\bibnamefont {Jach}},\
  }\href {\doibase 10.1103/PhysRevB.29.5617} {\bibfield  {journal} {\bibinfo
  {journal} {Phys. Rev. B}\ }\textbf {\bibinfo {volume} {29}},\ \bibinfo
  {pages} {5617} (\bibinfo {year} {1984})}\BibitemShut {NoStop}%
\bibitem [{\citenamefont {Zhang}\ \emph {et~al.}(2020)\citenamefont {Zhang},
  \citenamefont {Gao},\ and\ \citenamefont {Xiao}}]{zhang_topological_2020}%
  \BibitemOpen
  \bibfield  {author} {\bibinfo {author} {\bibfnamefont {Y.}~\bibnamefont
  {Zhang}}, \bibinfo {author} {\bibfnamefont {Y.}~\bibnamefont {Gao}}, \ and\
  \bibinfo {author} {\bibfnamefont {D.}~\bibnamefont {Xiao}},\ }\href {\doibase
  10.1103/PhysRevB.101.041410} {\bibfield  {journal} {\bibinfo  {journal}
  {Phys. Rev. B}\ }\textbf {\bibinfo {volume} {101}},\ \bibinfo {pages}
  {041410} (\bibinfo {year} {2020})},\ \bibinfo {note} {arXiv:1910.09001
  [cond-mat]}\BibitemShut {NoStop}%
\bibitem [{\citenamefont {Principi}\ \emph {et~al.}(2009)\citenamefont
  {Principi}, \citenamefont {Polini},\ and\ \citenamefont
  {Vignale}}]{principi_linear_2009}%
  \BibitemOpen
  \bibfield  {author} {\bibinfo {author} {\bibfnamefont {A.}~\bibnamefont
  {Principi}}, \bibinfo {author} {\bibfnamefont {M.}~\bibnamefont {Polini}}, \
  and\ \bibinfo {author} {\bibfnamefont {G.}~\bibnamefont {Vignale}},\ }\href
  {\doibase 10.1103/PhysRevB.80.075418} {\bibfield  {journal} {\bibinfo
  {journal} {Phys. Rev. B}\ }\textbf {\bibinfo {volume} {80}},\ \bibinfo
  {pages} {075418} (\bibinfo {year} {2009})}\BibitemShut {NoStop}%
\bibitem [{\citenamefont {Redlich}(1984{\natexlab{a}})}]{redlich_gauge_1984}%
  \BibitemOpen
  \bibfield  {author} {\bibinfo {author} {\bibfnamefont {A.~N.}\ \bibnamefont
  {Redlich}},\ }\href {\doibase 10.1103/PhysRevLett.52.18} {\bibfield
  {journal} {\bibinfo  {journal} {Phys. Rev. Lett.}\ }\textbf {\bibinfo
  {volume} {52}},\ \bibinfo {pages} {18} (\bibinfo {year}
  {1984}{\natexlab{a}})}\BibitemShut {NoStop}%
\bibitem [{\citenamefont {Redlich}(1984{\natexlab{b}})}]{redlich_parity_1984}%
  \BibitemOpen
  \bibfield  {author} {\bibinfo {author} {\bibfnamefont {A.~N.}\ \bibnamefont
  {Redlich}},\ }\href {\doibase 10.1103/PhysRevD.29.2366} {\bibfield  {journal}
  {\bibinfo  {journal} {Phys. Rev. D}\ }\textbf {\bibinfo {volume} {29}},\
  \bibinfo {pages} {2366} (\bibinfo {year} {1984}{\natexlab{b}})}\BibitemShut
  {NoStop}%
\bibitem [{\citenamefont {Koshino}\ and\ \citenamefont
  {Ando}(2011)}]{koshino_orbital_2011}%
  \BibitemOpen
  \bibfield  {author} {\bibinfo {author} {\bibfnamefont {M.}~\bibnamefont
  {Koshino}}\ and\ \bibinfo {author} {\bibfnamefont {T.}~\bibnamefont {Ando}},\
  }\href {\doibase 10.1088/1742-6596/334/1/012005} {\bibfield  {journal}
  {\bibinfo  {journal} {J. Phys.: Conf. Ser.}\ }\textbf {\bibinfo {volume}
  {334}},\ \bibinfo {pages} {012005} (\bibinfo {year} {2011})}\BibitemShut
  {NoStop}%
\bibitem [{\citenamefont {Wang}\ \emph {et~al.}(2017)\citenamefont {Wang},
  \citenamefont {Cooper}, \citenamefont {Halperin},\ and\ \citenamefont
  {Stern}}]{wang_particle-hole_2017}%
  \BibitemOpen
  \bibfield  {author} {\bibinfo {author} {\bibfnamefont {C.}~\bibnamefont
  {Wang}}, \bibinfo {author} {\bibfnamefont {N.~R.}\ \bibnamefont {Cooper}},
  \bibinfo {author} {\bibfnamefont {B.~I.}\ \bibnamefont {Halperin}}, \ and\
  \bibinfo {author} {\bibfnamefont {A.}~\bibnamefont {Stern}},\ }\href
  {\doibase 10.1103/PhysRevX.7.031029} {\bibfield  {journal} {\bibinfo
  {journal} {Phys. Rev. X}\ }\textbf {\bibinfo {volume} {7}},\ \bibinfo {pages}
  {031029} (\bibinfo {year} {2017})}\BibitemShut {NoStop}%
\bibitem [{\citenamefont {Ji}\ and\ \citenamefont
  {Shi}(2021)}]{ji_emergence_2021}%
  \BibitemOpen
  \bibfield  {author} {\bibinfo {author} {\bibfnamefont {G.}~\bibnamefont
  {Ji}}\ and\ \bibinfo {author} {\bibfnamefont {J.}~\bibnamefont {Shi}},\
  }\href {\doibase 10.1103/PhysRevResearch.3.043055} {\bibfield  {journal}
  {\bibinfo  {journal} {Phys. Rev. Research}\ }\textbf {\bibinfo {volume}
  {3}},\ \bibinfo {pages} {043055} (\bibinfo {year} {2021})}\BibitemShut
  {NoStop}%
\bibitem [{\citenamefont {Ji}\ and\ \citenamefont {Shi}(2020)}]{ji_berry_2020}%
  \BibitemOpen
  \bibfield  {author} {\bibinfo {author} {\bibfnamefont {G.}~\bibnamefont
  {Ji}}\ and\ \bibinfo {author} {\bibfnamefont {J.}~\bibnamefont {Shi}},\
  }\href {\doibase 10.1103/PhysRevResearch.2.033329} {\bibfield  {journal}
  {\bibinfo  {journal} {Phys. Rev. Research}\ }\textbf {\bibinfo {volume}
  {2}},\ \bibinfo {pages} {033329} (\bibinfo {year} {2020})}\BibitemShut
  {NoStop}%
\bibitem [{\citenamefont {Tőke}\ and\ \citenamefont
  {Jain}(2007)}]{toke_su4_2007}%
  \BibitemOpen
  \bibfield  {author} {\bibinfo {author} {\bibfnamefont {C.}~\bibnamefont
  {Tőke}}\ and\ \bibinfo {author} {\bibfnamefont {J.~K.}\ \bibnamefont
  {Jain}},\ }\href {\doibase 10.1103/PhysRevB.75.245440} {\bibfield  {journal}
  {\bibinfo  {journal} {Phys. Rev. B}\ }\textbf {\bibinfo {volume} {75}},\
  \bibinfo {pages} {245440} (\bibinfo {year} {2007})}\BibitemShut {NoStop}%
\bibitem [{\citenamefont {Thouless}\ \emph {et~al.}(1982)\citenamefont
  {Thouless}, \citenamefont {Kohmoto}, \citenamefont {Nightingale},\ and\
  \citenamefont {den Nijs}}]{thouless_quantized_1982}%
  \BibitemOpen
  \bibfield  {author} {\bibinfo {author} {\bibfnamefont {D.~J.}\ \bibnamefont
  {Thouless}}, \bibinfo {author} {\bibfnamefont {M.}~\bibnamefont {Kohmoto}},
  \bibinfo {author} {\bibfnamefont {M.~P.}\ \bibnamefont {Nightingale}}, \ and\
  \bibinfo {author} {\bibfnamefont {M.}~\bibnamefont {den Nijs}},\ }\href
  {\doibase 10.1103/PhysRevLett.49.405} {\bibfield  {journal} {\bibinfo
  {journal} {Phys. Rev. Lett.}\ }\textbf {\bibinfo {volume} {49}},\ \bibinfo
  {pages} {405} (\bibinfo {year} {1982})}\BibitemShut {NoStop}%
\bibitem [{\citenamefont {Schakel}(1991)}]{schakel_relativistic_1991}%
  \BibitemOpen
  \bibfield  {author} {\bibinfo {author} {\bibfnamefont {A.~M.~J.}\
  \bibnamefont {Schakel}},\ }\href {\doibase 10.1103/PhysRevD.43.1428}
  {\bibfield  {journal} {\bibinfo  {journal} {Phys. Rev. D}\ }\textbf {\bibinfo
  {volume} {43}},\ \bibinfo {pages} {1428} (\bibinfo {year}
  {1991})}\BibitemShut {NoStop}%
\bibitem [{\citenamefont {Mikitik}\ and\ \citenamefont
  {Sharlai}(1999)}]{mikitik_manifestation_1999}%
  \BibitemOpen
  \bibfield  {author} {\bibinfo {author} {\bibfnamefont {G.~P.}\ \bibnamefont
  {Mikitik}}\ and\ \bibinfo {author} {\bibfnamefont {Y.~V.}\ \bibnamefont
  {Sharlai}},\ }\href {\doibase 10.1103/PhysRevLett.82.2147} {\bibfield
  {journal} {\bibinfo  {journal} {Phys. Rev. Lett.}\ }\textbf {\bibinfo
  {volume} {82}},\ \bibinfo {pages} {2147} (\bibinfo {year}
  {1999})}\BibitemShut {NoStop}%
\bibitem [{\citenamefont {Lui}\ \emph {et~al.}(2011)\citenamefont {Lui},
  \citenamefont {Li}, \citenamefont {Mak}, \citenamefont {Cappelluti},\ and\
  \citenamefont {Heinz}}]{lui_observation_2011}%
  \BibitemOpen
  \bibfield  {author} {\bibinfo {author} {\bibfnamefont {C.~H.}\ \bibnamefont
  {Lui}}, \bibinfo {author} {\bibfnamefont {Z.}~\bibnamefont {Li}}, \bibinfo
  {author} {\bibfnamefont {K.~F.}\ \bibnamefont {Mak}}, \bibinfo {author}
  {\bibfnamefont {E.}~\bibnamefont {Cappelluti}}, \ and\ \bibinfo {author}
  {\bibfnamefont {T.~F.}\ \bibnamefont {Heinz}},\ }\href {\doibase
  10.1038/nphys2102} {\bibfield  {journal} {\bibinfo  {journal} {Nature Phys}\
  }\textbf {\bibinfo {volume} {7}},\ \bibinfo {pages} {944} (\bibinfo {year}
  {2011})}\BibitemShut {NoStop}%
\bibitem [{\citenamefont {Bao}\ \emph {et~al.}(2011)\citenamefont {Bao},
  \citenamefont {Jing}, \citenamefont {Velasco}, \citenamefont {Lee},
  \citenamefont {Liu}, \citenamefont {Tran}, \citenamefont {Standley},
  \citenamefont {Aykol}, \citenamefont {Cronin}, \citenamefont {Smirnov},
  \citenamefont {Koshino}, \citenamefont {McCann}, \citenamefont {Bockrath},\
  and\ \citenamefont {Lau}}]{bao_stacking-dependent_2011}%
  \BibitemOpen
  \bibfield  {author} {\bibinfo {author} {\bibfnamefont {W.}~\bibnamefont
  {Bao}}, \bibinfo {author} {\bibfnamefont {L.}~\bibnamefont {Jing}}, \bibinfo
  {author} {\bibfnamefont {J.}~\bibnamefont {Velasco}}, \bibinfo {author}
  {\bibfnamefont {Y.}~\bibnamefont {Lee}}, \bibinfo {author} {\bibfnamefont
  {G.}~\bibnamefont {Liu}}, \bibinfo {author} {\bibfnamefont {D.}~\bibnamefont
  {Tran}}, \bibinfo {author} {\bibfnamefont {B.}~\bibnamefont {Standley}},
  \bibinfo {author} {\bibfnamefont {M.}~\bibnamefont {Aykol}}, \bibinfo
  {author} {\bibfnamefont {S.~B.}\ \bibnamefont {Cronin}}, \bibinfo {author}
  {\bibfnamefont {D.}~\bibnamefont {Smirnov}}, \bibinfo {author} {\bibfnamefont
  {M.}~\bibnamefont {Koshino}}, \bibinfo {author} {\bibfnamefont
  {E.}~\bibnamefont {McCann}}, \bibinfo {author} {\bibfnamefont
  {M.}~\bibnamefont {Bockrath}}, \ and\ \bibinfo {author} {\bibfnamefont
  {C.~N.}\ \bibnamefont {Lau}},\ }\href {\doibase 10.1038/nphys2103} {\bibfield
   {journal} {\bibinfo  {journal} {Nature Phys}\ }\textbf {\bibinfo {volume}
  {7}},\ \bibinfo {pages} {948} (\bibinfo {year} {2011})}\BibitemShut {NoStop}%
\bibitem [{\citenamefont {Zhou}\ \emph {et~al.}(2007)\citenamefont {Zhou},
  \citenamefont {Gweon}, \citenamefont {Fedorov}, \citenamefont {First},
  \citenamefont {de~Heer}, \citenamefont {Lee}, \citenamefont {Guinea},
  \citenamefont {Castro~Neto},\ and\ \citenamefont
  {Lanzara}}]{zhou_substrate-induced_2007}%
  \BibitemOpen
  \bibfield  {author} {\bibinfo {author} {\bibfnamefont {S.~Y.}\ \bibnamefont
  {Zhou}}, \bibinfo {author} {\bibfnamefont {G.-H.}\ \bibnamefont {Gweon}},
  \bibinfo {author} {\bibfnamefont {A.~V.}\ \bibnamefont {Fedorov}}, \bibinfo
  {author} {\bibfnamefont {P.~N.}\ \bibnamefont {First}}, \bibinfo {author}
  {\bibfnamefont {W.~A.}\ \bibnamefont {de~Heer}}, \bibinfo {author}
  {\bibfnamefont {D.-H.}\ \bibnamefont {Lee}}, \bibinfo {author} {\bibfnamefont
  {F.}~\bibnamefont {Guinea}}, \bibinfo {author} {\bibfnamefont {A.~H.}\
  \bibnamefont {Castro~Neto}}, \ and\ \bibinfo {author} {\bibfnamefont
  {A.}~\bibnamefont {Lanzara}},\ }\href {\doibase 10.1038/nmat2003} {\bibfield
  {journal} {\bibinfo  {journal} {Nature Mater}\ }\textbf {\bibinfo {volume}
  {6}},\ \bibinfo {pages} {770} (\bibinfo {year} {2007})}\BibitemShut {NoStop}%
\bibitem [{\citenamefont {Choi}\ \emph {et~al.}(2010)\citenamefont {Choi},
  \citenamefont {Jhi},\ and\ \citenamefont {Son}}]{choi_controlling_2010}%
  \BibitemOpen
  \bibfield  {author} {\bibinfo {author} {\bibfnamefont {S.-M.}\ \bibnamefont
  {Choi}}, \bibinfo {author} {\bibfnamefont {S.-H.}\ \bibnamefont {Jhi}}, \
  and\ \bibinfo {author} {\bibfnamefont {Y.-W.}\ \bibnamefont {Son}},\ }\href
  {\doibase 10.1021/nl101617x} {\bibfield  {journal} {\bibinfo  {journal} {Nano
  Lett.}\ }\textbf {\bibinfo {volume} {10}},\ \bibinfo {pages} {3486} (\bibinfo
  {year} {2010})}\BibitemShut {NoStop}%
\end{thebibliography}%

\end{document}